\documentclass[11pt,reqno]{amsart}
\usepackage{amsmath,amsthm,amssymb,bm,bbm,dsfont}
\usepackage{mathtools}\mathtoolsset{centercolon}
\usepackage[shortlabels]{enumitem}
\usepackage[mathscr]{eucal}
\usepackage{amsaddr}
\usepackage{upgreek}
\usepackage[left=2cm,right=2cm,top=2cm,bottom=2cm]{geometry}
\mathtoolsset{showonlyrefs=true}
\usepackage{setspace}
\usepackage{graphicx}
\usepackage{verbatim}
\usepackage{float}
\usepackage{placeins}
\usepackage{array, soul}
\usepackage{booktabs}
\usepackage{threeparttable}
\usepackage[update,prepend]{epstopdf}
\usepackage{multirow}
\usepackage{amsfonts,amssymb,dsfont}
\usepackage[abs]{overpic}
\usepackage{xfrac}

\usepackage[usenames,dvipsnames]{color}
\usepackage{colortbl}
\usepackage[hidelinks]{hyperref}
\hypersetup{
	unicode=false,          
	pdftoolbar=true,        
	pdfmenubar=true,        
	pdffitwindow=false,     
	pdfstartview={FitH},    
	pdftitle={My title},    
	pdfauthor={Author},     
	pdfsubject={Subject},   
	pdfcreator={Creator},   
	pdfproducer={Producer}, 
	pdfkeywords={keyword1} {key2} {key3}, 
	pdfnewwindow=true,      
	colorlinks=true,        
	linkcolor=Red,          
	citecolor=ForestGreen,  
	filecolor=Magenta,      
	urlcolor=BlueViolet,    
}
\usepackage{doi}
\usepackage{url}
\usepackage{caption, subcaption}
\usepackage{enumitem}

\makeatletter
\ifx\@NODS\undefined%

\let\mathbb=\mathds
\else%
\fi
\makeatother

\DeclareMathOperator*{\argmin}{\arg\min}

\DeclareMathOperator{\Tr}{Tr}

\DeclareMathOperator{\F}{\mathcal{F}}
\DeclareMathOperator{\M}{\mathbb{M}}

\newcommand{\pl}{\hspace{.1cm}}
\newcommand{\ten}{\otimes}
\newcommand{\al}{\alpha}
\newcommand{\si}{\sigma}
\newcommand{\norm}[2]{\parallel \! #1 \! \parallel_{#2}}

\newcommand{\iS}{{\mathcal{S}}}
\newcommand{\iM}{{\mathcal{M}}}
\newcommand{\iN}{{\mathcal{N}}}

\newcommand{\be}{{\mathbf e}}

\def\0{{\mathbf{0}}}
\def\1{{\mathbf{1}}}
\def\2{{\mathbf{2}}}
\def\3{{\mathbf{3}}}
\def\4{{\mathbf{4}}}
\def\5{{\mathbf{5}}}
\def\6{{\mathbf{6}}}

\def\7{{\mathbf{7}}}
\def\8{{\mathbf{8}}}
\def\9{{\mathbf{9}}}

\def\be{\begin{equation}}
\def\ee{\end{equation}}
\def\bea{\begin{eqnarray}}
\def\eea{\end{eqnarray}}

\def\eps{\varepsilon}

\def\M{{\mathcal M}}

\theoremstyle{plain}
\newtheorem{theo}{Theorem} 
\newtheorem{prop}[theo]{Proposition} 
\newtheorem{lemm}[theo]{Lemma} 
\newtheorem{coro}[theo]{Corollary} 

\newtheorem*{prop2}{Proposition~\ref{prop:I2_R}}
\newtheorem*{prop3}{Proposition~\ref{prop:I2}}

\theoremstyle{definition}
\newtheorem{defn}[theo]{Definition} 

\theoremstyle{remark}
\newtheorem{remark}{Remark}[section]

\newcommand{\vertiii}[1]{{\left\vert\kern-0.25ex\left\vert\kern-0.25ex\left\vert #1
		\right\vert\kern-0.25ex\right\vert\kern-0.25ex\right\vert}}

\makeatletter
\newcommand{\opnorm}{\@ifstar\@opnorms\@opnorm}
\newcommand{\@opnorms}[1]{%
	\left|\mkern-1.5mu\left|\mkern-1.5mu\left|
	#1
	\right|\mkern-1.5mu\right|\mkern-1.5mu\right|
}
\newcommand{\@opnorm}[2][]{%
	\mathopen{#1|\mkern-1.5mu#1|\mkern-1.5mu#1|}
	#2
	\mathclose{#1|\mkern-1.5mu#1|\mkern-1.5mu#1|}
}
\makeatother

\begin{document}

\let\origmaketitle\maketitle
\def\maketitle{
	\begingroup
	\def\uppercasenonmath##1{} 
	\let\MakeUppercase\relax 
	\origmaketitle
	\endgroup
}

\title{\bfseries \Large{Properties of Noncommutative R\'enyi and Augustin Information
		}}

\author{ \large \textsc{Hao-Chung Cheng$^{1,2,3}$, Li Gao$^{4}$ and Min-Hsiu Hsieh$^3$}}
\address{\small  	
$^{1}$Department of Electrical Engineering \& Graduate Institute of Communication Engineering\\
$^{2}$Department of Mathematics \& Institute of Applied Mathematical Sciences\\
National Taiwan University, 106 Taipei, Taiwan (R.O.C.)\\
$^{3}$Hon Hai (Foxconn) Quantum Computing Centre, New Taipei City 236, Taiwan (R.O.C.)\\
$^{4}$Department of Mathematics, University of Houston, TX 77004, United States
}

\email{\href{mailto:haochung.ch@gmail.com}{haochung.ch@gmail.com}}
\email{\href{mailto:gaolimath@gmail.com}{gaolimath@gmail.com}}
\email{\href{mailto:minhsiuh@gmail.com}{minhsiuh@gmail.com}}

\date{\today}
\begin{abstract} 
R\'enyi and Augustin information are generalizations of mutual information defined via the R\'enyi divergence, playing a significant role in evaluating the performance of information processing tasks by virtue of its connection to the error exponent analysis. In quantum information theory, there are three generalizations of the classical R\'enyi divergence---the Petz's, sandwiched, and log-Euclidean versions, that possess meaningful operational interpretation. However, the associated quantum R\'enyi and Augustin information are much less explored compared with their classical counterpart, and lacking crucial properties hinders applications of these quantities to error exponent analysis in the quantum regime.

The goal of this paper is to analyze fundamental properties of the R\'enyi and Augustin information from a noncommutative measure-theoretic perspective.  Firstly, we prove the uniform equicontinuity for all three quantum versions of R\'enyi and Augustin information, and it hence  yields the joint continuity of these quantities in order and prior input distributions. 
Secondly, we establish the concavity of the scaled R\'enyi and Augustin information in the region of $s\in(-1,0)$ for both Petz's and the sandwiched versions. This completes the open questions raised by Holevo [\href{https://ieeexplore.ieee.org/document/868501/}{\textit{IEEE Trans.~Inf.~Theory}, \textbf{46}(6):2256--2261, 2000}], and Mosonyi and Ogawa [\href{https://doi.org/10.1007/s00220-017-2928-4/}{\textit{Commun.~Math.~Phys.}, \textbf{355}(1):373--426, 2017}]. 
For the applications, we show that the strong converse exponent in classical-quantum channel coding satisfies a minimax identity, which means that the strong converse exponent can be attained by the best constant composition code.
The established concavity is further employed to prove an entropic duality between classical data compression with quantum side information and classical-quantum channel coding, and a Fenchel duality in joint source-channel coding with quantum side information.

\end{abstract}
\maketitle


\section{Introduction} \label{sec:introduction}

Error exponent analysis aims at evaluating the exponential behavior of the performance (e.g.~the error probability or success probability) of the underlying system when certain \emph{size} or \emph{rate} is fixed.
Early studies can be found in hypothesis testing, detection and estimation theory, and varieties of statistical applications \cite{Che52, Che56, Hoe63, Hoe67, Leh86, Bah67, Bah71, Bah83, KK86, Cha80, Ber71, Ber80, DZ98}.
It is arguably a substantial research topic in information theory because the analysis can be viewed as a refinement of Shannon's seminal source coding and channel coding theorem~\cite{Sha48}. In this paper, we focus on the problems of information transmission or the so-called channel coding.
Let $\mathscr{W} : x \mapsto W_x$ be a probabilistic channel that maps symbols from the input alphabet $\mathcal{X}$ to an measurable output space. The goal of a communication system is to design a good coding strategy for $n$ uses of the channel that minimizes the error probability of decoding. Drawing a connection to the large deviation principle \cite{DZ98},  the optimal exponent given a fixed transmission rate $R$ is determined by the \emph{Fenchel-Legendre transform} of the scaled R\'enyi information $E_0^{\texttt{r}}(s,P)$ (maximized over all priors $P$) \cite{Fei55, Sha59b, Fan61, Gal65, Gal68, SGB67, SGB67b, Gal68, Ari73, DK79, PV10b, Nak16b}:\footnote{More precisely, Eq.~\eqref{eq:exponent_err} was proven for any fixed rate below the channel capacity and above the critical rate \cite{Gal65, SGB67, Gal68, CK11}, the rate at which the slope of the right-hand side of \eqref{eq:exponent_err} is $-1$.
	Recently, Nakibo\u{g}lu in \cite[Lemma 29]{Nak16b} showed that Eq.~\eqref{eq:exponent_err} holds for any fixed rate greater than $C_{\frac{1}{1+L}, \mathscr{W}}$ under list decoding \cite{CK11} with list size $L \in \{1,\ldots, M-1\}$, where $M$ is the size of the message set.
}
\begin{align}
\lim_{n\to \infty} - \frac1n \log \eps^\star (n,R) &=  \sup_{ 0\leq s \leq 0} \sup_{P} \left\{ E_0^{\texttt{r}}(s,P) - sR  \right\}, \quad R< C_{\mathscr{W}} \label{eq:exponent_err} \\
\lim_{n\to \infty} - \frac1n \log \left[ 1 - \eps^\star (n,R) \right] &= \sup_{-1< s < 0} \inf_P \left\{ E_0^{\texttt{r}}(s,P) - sR  \right\}, \quad R>C_{\mathscr{W}} \label{eq:exponent_sc}
\end{align}
where $\eps^\star(n,R)$ denotes the optimal error probability; $C_{\mathscr{W}}$ is the channel capacity; and $E_0^{\texttt{r}}(s,P)$ for a prior probability mass function $P$ is called the \emph{auxiliary function} introduced by Gallager \cite{Gal65, Gal68}:
\begin{align} \label{eq:Gallager}
E_0^{\texttt{r}}(s,P) := -\log \int \left( \sum_x P(x) \left(\frac{\mathrm{d} W_x}{ \mathrm{d} \nu  }\right)^{\frac{1}{1+s}} \right)^{1+s} \mathrm{d}\nu, \quad s>-1,
\end{align}
where $\nu$ is any reference measure\footnote{Note that the quantity $E_0^{\texttt{r}}(s,P)$  does not depend on the choice of the reference measure.} such that $W_x$ is absolutely continuous with respect to $\nu$ for all $x$ with $P(x)>0$.

The auxiliary function presented above has a close relation to a one-parameter generalization of Shannon's mutual information.
Sibson \cite{Sib69} introduced one candidate in terms of R\'enyi's divergence $D_\alpha$ \cite{Ren61, EH14} and showed that it equals a scaled version of Gallager's auxiliary function:
\begin{align}
I_\alpha^{\texttt{r}}(P,\mathscr{W}) &:= \inf_{q} D_\alpha\left( P\circ \mathscr{W} \| P\otimes q \right) \label{eq:information_Renyi} \\
&= \left.\frac{E_0^{\texttt{r}}(s,P)}{s}\right|_{s = \frac{1-\alpha}{\alpha} }, \label{eq:Sibson_classical}
\end{align}
where the infimum is taken over all probability measures $q$ on the output measurable space, and $P\circ \mathscr{W}$ denotes the joint probability measure on the product of input and output spaces.
We termed $I_\alpha^{\texttt{r}}(P,\mathscr{W})$ \emph{the order-$\alpha$ R\'enyi information for a prior $P$}.

Augustin \cite{Aug78} and Csisz{\'a}r \cite{Csi95} proposed another generalization\footnote{There is another version defined by Arimoto \cite{Ari77}. After maximizing over all priors, the three quantities correspond to the order-$\alpha$ R\'enyi capacity in Eq.~\eqref{eq:radius0}. However, we omit the discussion of this version due to its limited uses. We refer the readers' to the comparison by Verd\'u \cite{Ver15}
	and by Aishwarya and Madiman\cite{AM20}.	
} of Shannon's mutual information, which is termed as \emph{the order-$\alpha$ Augustin information for a prior $P$} \cite{Aug78, Csi95, Dal17, Nak18a}:
\begin{align} \label{eq:information_Aug}
I_\alpha^{\texttt{a}}(P,\mathscr{W}) &:= \inf_{q} \sum_{x} P(x) D_\alpha\left( W_x \| q \right).
\end{align}
When maximizing over all priors $P$, both R\'enyi information and Augustin information equal \emph{the order-$\alpha$ R\'enyi capacity} \cite{Kem74, Csi95, Nak16a, Nak18a}:
\begin{align} \label{eq:radius0}
C_{\alpha,\mathscr{W}} := \sup_P I_\alpha^{\texttt{r}}(P,\mathscr{W}) = \sup_P I_\alpha^{\texttt{a}} (P,\mathscr{W}).
\end{align}
One can define the auxiliary function associated with the Augustin information by drawing inspiration from Eq.~\eqref{eq:Sibson_classical}:
\begin{align}
E_0^{\texttt{a}}(s,P) := s I_{\frac{1}{1+s}}^{\texttt{a}}(P,\mathscr{W}).
\end{align}
Similar to the role of $E_0^{\texttt{r}}(s,P)$ in Eqs.~\eqref{eq:exponent_err} and \eqref{eq:exponent_sc}, the Fenchel-Legendre transform of $E_0^{\texttt{a}}(s,P)$ was shown to be equal to the optimal exponent of channel coding with constant composition codes
\cite{Bla74, Bla87, Aug66, Aug69, Aug78, Har68, HHH07, Gal94,
	DK79, CK81, Csi95, Csi98, CK11, SMF14, Sca14, Nak18b}.

In addition to the channel coding problems, the connections of the auxiliary functions\footnote{The auxiliary functions in different protocols are defined in a slightly different but similar way \cite{Hay07, CHT17, CH17, CHDH-2018, CHDH2-2018, CHDH3-2018}. We refer the readers to Section~\ref{sec:applications} for further discussion.} to the exponents in other information tasks, e.g.~source coding  and channel coding networks, have been established as well \cite{SW73, Gal76, Kos77, CK80, CK81, Csi82, Ahl80, AD82, Ahl14}.
This justifies the operational significance of the auxiliary functions in information theory.
Therefore, understanding their properties is of substantial interest and allows us to better characterize the performance of the information tasks.
Early works on the continuity properties were done by Gallager \cite[p.~28]{Gal65}, Shannon, Gallager, and Berlekamp \cite[p.~101]{SGB67}, and  Csisz\'{a}r and K{\"o}rner \cite{Csi67, CK11}. The first-order and second-order derivatives at $s=0$ correspond, respectively, to Shannon's mutual information and information variance \cite[p.~142]{Gal68}, \cite[Lemma 1]{AW14b}. Those properties are critical to high-order analysis in the finite blocklength regime \cite{PPV10, Tan14, SMG14, Sca14, AW14d} and moderate deviation analysis \cite{AW14b, CH17}.
The concavity of $E_0^{\texttt{r}}(s,P)$ in $s>-1$ was first proved by Gallager \cite[Theorem 5.6.3]{Gal68} using H\"older's inequality. Essentially, the concavity of $E_0^{\texttt{r}}$ is equivalent to Littlewood's version of H\"older's inequality\footnote{There are several versions of H\"older's inequality. The one used by Gallager \cite[(5B.10)]{Gal68} is $\sum_{j} a_j b_j \leq (\sum_j a_j^{1/(1-\theta)} )^{1-\theta} (\sum_j b_j^{\theta} )^{\theta}$ for all $a_j,b_j\geq 0$ and $\theta \in [0,1]$. On the other hand, Littlewood's version, which is also called interpolation inequality, states that $\|\mathbf{u}\|_{1/((1-\theta) p + \theta q )} \leq \|\mathbf{u}\|_{1/p}^{1-\theta} \|\mathbf{u}\|_{1/q}^{\theta}$, where $\|\mathbf{u}\|_p := (\sum_j w_j u_j^p)^{1/p}$ is the $p$-norm for nonnegative $(w_j)_j$.} \cite[Theorem 5.5.1]{Gar07}.
As for $E_0^{\texttt{a}}(s,P)$, Csisz\'{a}r  \cite[(A24), (A27)]{Csi95} (see also \cite[Theorem 30]{EH14},  \cite{BTC88}) proved a variational representation for the case of finite-dimensional output spaces:
\begin{align} \label{eq:information_var}
E_0^{\texttt{a}}(s,P) = \inf_{\mathscr{V}} \left\{ \sum_x P(x) D\left( V_x\|W_x\right) + s I(P,\mathscr{V})  \right\}, \quad s > -1,
\end{align}
where the infimum is taken over all dummy channels on the same input and output spaces of $\mathscr{W}$; $I(P,\mathscr{V})$ is Shannon's mutual information; and $D(\cdot\|\cdot)$ is the Kullback--Leibler divergence.
Then, the concavity in $s>-1$ immediately follows because a pointwise infimum of linear functions is concave.
We remark that the concavity property in $s$ has numerous usefulness. For example, it determines the convexity and decreases of the entropic quantities in $R$ \cite[p.~142]{Gal68}, and it is indispensable in proving the saddle-point property in sphere-packing exponents \cite{AW14, CHT17, CHDH-2018}, and the moderate deviations \cite{AW14b}.
The properties of the auxiliary functions can also be derived via those of the R\'enyi and Augustin information. We refer the readers to the review literature by Ho and Verd{\'u} \cite{HV15, Ver15}, Dalai \cite{Dal17}, and the excellent expositions by Nakibo\u{g}lu \cite{Nak16a, Nak18a} from a measure-theoretic aspect.

In classical information theory, the (channel) output space consists of probability measures. It can be extended to more general noncommutative measure spaces, i.e.~von Neumann algebras, as any quantum-mechanical system can be modeled by a density operator. One prominent example is the classical-quantum channel coding, where the output space contains density matrices \cite{NC09, Hay06, Wilde2, Tom16}.
Therefore, one of the main aims of the current paper is to investigate the properties of the auxiliary functions using noncommutative $L_p$-theory. Moreover, the established results could be employed to perform refined analysis in quantum information processing tasks \cite{CHT17, CH17, CHDH-2018, CHDH2-2018, CHDH3-2018, Hao-Chung}.

There are at least three quantum generalizations of the classical R\'enyi divergence \cite{Ren61}: Petz's R\'enyi divergence $D_\alpha$ \cite{Pet86}, the sandwiched R\'enyi divergence $D_\alpha^*$ \cite{MDS+13,WWY14, BST18, Jen18a}, and the log-Euclidean R\'enyi divergence $D_\alpha^\flat$ \cite{ON00, MO17}. The quantum auxiliary functions are defined accordingly: for $(t) = \{\}$, $*$ or $\flat$,
\begin{align}
&E_0^{\texttt{r},(t)}(s,P) := s I_{\frac{1}{1+s}}^{\texttt{r},(t)} (P, \mathscr{W}),\\
&E_0^{\texttt{a},(t)}(s,P) := s I_{\frac{1}{1+s}}^{\texttt{a},(t)} (P, \mathscr{W}).
\end{align}
Due to the noncommutative nature, it is generally more difficult to derive properties for them. Furthermore, there are no closed-form expressions except for $E_0^{\texttt{r}}(s,P)$.
Actually, the three versions inherit different properties of the classical function. A quantum Sibson's identity holds for the Petz's version as in Eq.~\eqref{eq:Sibson_classical} \cite{SW12}; the sandwiched version relates to weighted noncommutative $L_p$-norms; and the log-Euclidean version satisfies the variational representation as in Eq.~\eqref{eq:information_var}. We will exploit these facts in our derivations later.

Burnashev and Holevo \cite{BH98, Hol00} first generalized Gallager's expression in Eq.~\eqref{eq:Gallager} to a quantum auxiliary function:
\begin{align} \label{eq:E0_0}
E_{0}^{\texttt{r}}(s,P) := - \log \Tr\left[ \left( \sum_{x} P(x) W_x^{\frac{1}{1+s}} \right)^{1+s} \right],
\end{align}
where the $\left\{W_x \right\}_x$ is a set of density operators in the output space.
Sharma and Warsi \cite{SW12} proved a quantum Sibson's identity to show that
the expressions in Eqs.~\eqref{eq:information_Renyi} and \eqref{eq:Sibson_classical} are equal to Petz's version $E_0^{\texttt{r}}(s,P)$.
If the density operators are all rank-one (i.e.~pure-state channels), Burnashev and Holevo \cite{BH98, Hol00} proved a random coding bound (i.e.~achievability) on the optimal error probability in terms of the Fenchel-Legendre transform of $E_0^{\texttt{r}}(s,P)$.
Burnashev and Holevo \cite{BH98, Hol00} also conjectured that their result holds when the output space consists of general density operators.
Hayashi proved an achievability bound with a sub-optimal auxiliary function \cite{HN03, Hay07}. Recently, Qi \text{et al.} extended Hayashi's expression to entanglement-assisted classical communications over quantum channels \cite{QWW18}.
The sphere-packing bound (i.e.~optimality) was first studied by Winter \cite{Win99}, and he proved the bound with the log-Euclidean version.
Recently, Dalai \cite{Dal13} and part of the present authors \cite{CHT17} established a sphere-packing bound for all codes with Petz's version when maximizing over all priors $P$ as in the Eq.~\eqref{eq:exponent_err}. The sphere-packing bound for constant composition codes was also proved by using $E_0^{\texttt{a}}(s,P)$ \cite{DW14, CHT17}.
Compared with Winter's result, Petz's version is tighter than the log-Euclidean when $R<C_{\mathscr{W}}$ by Golden-Thompson's inequality \cite{Gol65, Sym65, Tho65, Dal17, CHT17}.
In the strong converse regime ($R>C_\mathscr{W}$), Mosonyi and Ogawa \cite{MO17} proved that the strong converse exponent is determined by the sandwiched version, see Eq.~\eqref{eq:exponent_sc} with $E_0^{\texttt{r},*}(s,P)$.

Regarding the properties of the auxiliary functions, Holevo \cite{Hol00} conjectured that $E_0^{\texttt{r}}(s,P)$ is concave as in the classical case. Later, Fujii and Yanagi proved the concavity in the region $s\in[0,1]$ by directly analyzing the second-order derivatives. Part of the authors \cite{CH16} employed a technique---the concavity of matrix geometric means---to show the concavity for all $s\geq 0$. Mosonyi and Ogawa in \cite[Theorem 3.6, Lemma 5.13]{MO17} showed that the log-Euclidean version satisfies the variational representation as in Eq.~\eqref{eq:information_var}, the concavity of $E_0^{\texttt{r},\flat}(s,P)$ and $E_0^{\texttt{a},\flat}(s,P)$ on $s>-1$ thus holds \cite[Proposition B.5]{MO17}.
Most importantly, Mosonyi and Ogawa \cite[Proposition B.1]{MO17} showed that
\begin{align}
\alpha \mapsto (\alpha-1) D_\alpha^{(t)}\text{ is convex on } (0,1) \;\text{ implies that } \; s\mapsto E_0^{\texttt{r},(t)}(s,P) \text{ or } E_0^{\texttt{a},(t)}(s,P) \text{ is concave on } (0,\infty).
\end{align}
Since the convexity assumption is true by \cite[Lemma 3.12]{MO17}, the concavity on $s\geq 0$ for all the versions was proved. However, the concavity for Petz's versions $E_0^{\texttt{r}}(s,P)$ and $E_0^{\texttt{a}}(s,P)$, and the sandwiched versions $E_0^{\texttt{r},*}(s,P)$ and $E_0^{\texttt{a},*}(s,P)$ on $s\in(-1,0)$ remains unknown.

The main contribution of this paper is proving the continuity of the quantum auxiliary functions and completing the last missing part of the concavity.
First, we show that the finiteness of the order-$\alpha$ R\'enyi capacity implies the uniform equicontinuity of the R\'enyi information and Augustin information in prior, respectively, in the region $(0,\min\{1,\alpha\}]$ for Petz's version, in $(0,\alpha]$ for log-Euclidean version, and in $[1/2, \alpha]$ for sandwiched version (Propositions~\ref{prop:I2_R} and \ref{prop:I2}).
Combining with the continuity of the R\'enyi information and Augustin information in their orders, we thus prove the joint continuity of the auxiliary functions in the argument (Theorem~\ref{theo:joint_cont}).
Second, we establish the concavity property of the auxiliary functions on $s\in(-1,0)$ for both the Petz's and sandwiched versions (Theorems~\ref{theo:concave_*} and \ref{theo:concave_Petz}), which solves the open problems  raised by Holevo \cite{Hol00}, Mosonyi and Ogawa \cite{MO17}. Moreover, the concavity results hold for the densities from any finite von Neumann algebras.

In order to prove the concavity, our main technique is the complex interpolation for noncommutative $L_p$ spaces. Firstly, we show that the R\'enyi auxiliary function for the sandwiched version, i.e.~$E_0^{\texttt{r},*}(s,P)$ is related to the amalgamated $L_p$-norm introduced by Junge and Parcet \cite{JP10} (see also the vector-valued $L_p$ space norm by Pisier \cite{Pis98}). Its concavity in $s\in(-1,0)$ can be derived from complex interpolation of the amalgamated $L_1^p(\iN\subset \iM)$-norm that for a positive $\rho \in\mathcal{M}$,
\begin{align}
\norm{\rho}{L_1^p(\iN\subset\iM)}=
\inf_{\si \in \iS(\iN)} \left\| \si^{\frac{1-p}{2p}}\rho \si^{\frac{1-p}{2p}} \right\|_p, \quad p \in [1,\infty), \;
\label{eq:augmented}
\end{align}
Here $\mathcal{M}$ is a semifinite von Neumann algebra\footnote{The readers not familiar with von Neumann algebras can think $\iM=\mathcal{B(H)}$ the bounded operators on a Hilbert space}; $\mathcal{S(\mathcal{N})}$ denotes all density operators in the subalgebra $\mathcal{N}\subset\mathcal{M}$; and $\|\cdot\|_p$ is the $L_p$-norm on $\mathcal{M}$.
The interpolation relation was proved in \cite{JP10}\footnote{See Eq.~\eqref{eq:inter_sandwiched} in Section~\ref{sec:interpolation} for the more detailed expression.}.
Secondly, the concavity of the Augustin auxiliary function for the sandwiched version follows from an interpolation type inequality: the log-convexity of the map
\begin{align}
\frac{1}{p} \mapsto
\inf_{\si \in \iS(\mathcal{H})} \left\| (\si^{\otimes n})^{\frac{1-p}{2p}}\rho (\si^{\otimes n})^{\frac{1-p}{2p}} \right\|_p, \quad p \in [1,\infty), \; \rho\in \mathcal{B}(\mathcal{H})^{\otimes n}.
\label{eq:augmented2}
\end{align}
Here the infimum is no longer taken over all density operators in a subalgebra $\mathcal{N}$ as in Eq.~\eqref{eq:augmented} but over all tensor-product states. This interpolation type inequality, i.e.~the log-convexity given in \eqref{eq:augmented2}, is shown in Theorem~\ref{theo:concave_*}.

Regarding the auxiliary functions of Petz form,
for $E_0^{\texttt{r}}(s,P)$ we require the log-convexity of the map
\begin{align}
\frac{1}{p}\mapsto \inf_{\sigma\in \iS(\mathcal{H})} \left( \Tr\left[  \left(\sum_{x\in\mathcal{X}} P(x) W_x^p\right) \sigma^{1-p} \right] \right)^{\frac1p}, \quad p \in [1,\infty).
\end{align}
This quantity is related to a new noncommutative Sibson's identity (Proposition~\ref{prop:Sibson}) and shows that the R\'enyi auxiliary function of Petz's form admits a representation
$
\Tr[ ( \mathbb{E}(\rho^p) )^{1/p} ]
$, where $\mathbb{E}:\mathcal{M}\to \mathcal{N}$ is the conditional expectation.
The concavity of $E_0^\texttt{r}(s,P)$ can be derived from the log-convexity of the map (Proposition~\ref{theo:inter_Petz})
\begin{align}
\frac{1}{p} \mapsto \Tr\left[ \left( \mathbb{E}(\rho^p) \right)^{\frac{1}{p}} \right], \quad p \in [1,\infty),\pl \rho\in\mathcal{M},
\end{align}
For the Augustin auxiliary function of Petz's form $E_0^\texttt{a}(s,P)$, we prove the log-convexity of the map
\begin{align}
\frac{1}{p} \mapsto
\inf_{\si \in \iS(\mathcal{H})} \left(\Tr\left[ \rho^p (\si^{\otimes n})^{{1-p}} \right]\right)^{\frac1p}, \quad p \in [1,\infty), \; \rho\in \mathcal{B}(\mathcal{H})^{\otimes n}.
\label{eq:augmented3}
\end{align}

The possible and future applications of the established results are the following. The joint continuity is useful in higher-order analysis in finite blocklength regime, and the variable-length classical data compression with quantum side information (also called the classical-quantum  Slepian-Wolf source coding) \cite{CHDH2-2018}.
For $s\in (-1,0)$, the auxiliary functions $E_0^{\texttt{r},(t)}(s,P)$ and $E_0^{\texttt{a},(t)}(s,P)$ are quasi-convex in prior $P$ by  Propositions~\ref{prop:I2_R} and~\ref{prop:I2}.
Hence, the established concavity in $s\in(-1,0)$ together with Sion's minimax theorem \cite{Sio58} immediately implies a minimax identity for the strong converse exponent:
\begin{align}
\sup_{-1< s < 0} \inf_P \left\{ E_0^{\texttt{r},*}(s,P) - sR  \right\} &=
\inf_P \sup_{-1< s < 0}  \left\{ E_0^{\texttt{r},*}(s,P) - sR  \right\} \\
= \sup_{-1< s < 0} \inf_P \left\{ E_0^{\texttt{a},*}(s,P) - sR  \right\} &=
\inf_P \sup_{-1< s < 0}  \left\{ E_0^{\texttt{a},*}(s,P) - sR  \right\}.
\end{align}
Moreover, the concavity is critical in proving an entropic duality between the classical data compression with quantum side information and a classical-quantum channel coding \cite{CHDH2-2018}, and a Fenchel duality in joint source-channel coding with quantum side information \cite{CHDH3-2018}.
We provide the comparisons of different notions of the auxiliary functions in Table~\ref{table:comparison}.

The paper is organized as follows. In Section~\ref{sec:preliminaries}, we introduce the definition and notation for various quantum entropic quantities and the corresponding auxiliary functions.
Section~\ref{sec:interpolation} reviews the basics of complex interpolation and proves an interpolation inequality.
We prove several properties of the auxiliary functions in Section~\ref{sec:properties}.
In Section~\ref{sec:applications}, we discuss their applications in quantum information theory. Finally, we conclude this paper in Section~\ref{sec:conclusions}.


\begin{table}[th!]
	\centering
	\resizebox{1\columnwidth}{!}{
		\begin{tabular}{@{}>{\columncolor[gray]{0.90}}cccccccc@{}}	
			\toprule
			
			Setting & Range of $s$ & Positivity & Monotone & Concave in $s$ & Continuity & $\left.\frac{\partial}{\partial s}\right|_{s=0}$ & $-\left.\frac{\partial^2}{\partial s^2}\right|_{s=0}$ \\
			
			\midrule
			\midrule		
			
			& $  [-1,0) $ & $<0$ & \multirow{2}{*}{$\nearrow$} & \multirow{2}{*}{$\cap$} & \multirow{2}{*}{$\checkmark$} & \multirow{2}{*}{$I(P,\mathscr{W})$} & \multirow{2}{*}{$U^{(t)}(P,\mathscr{W})$} \\
			\multirow{-2}{*}{$E_{0}^{\texttt{r},(t)}(s,P)$} & $[0,\infty]$ & $\geq0$ & & &  & & \\
			
			\midrule
			
			& $[-1,0)$ & $<0$ & \multirow{2}{*}{$\nearrow$} & \multirow{2}{*}{$\cap$} & \multirow{2}{*}{$\checkmark$} & \multirow{2}{*}{$I(P,\mathscr{W})$} & \multirow{2}{*}{$V^{(t)}(P,\mathscr{W})$} \\
			\multirow{-2}{*}{$E_{0}^{\texttt{a},(t)}(s,P)$} & $  [0,\infty] $ & $\geq0$ & &  & & & \\
			
			\midrule
			
			& $ [-1,0)$ & $\geq0$ & \multirow{2}{*}{$\searrow$} & \multirow{2}{*}{$\cap$} & \multirow{2}{*}{$\checkmark$}  & \multirow{2}{*}{$-H(X|B)_\rho$} & \multirow{2}{*}{$V^{(t)}(X|B)_\rho$} \\
			\multirow{-2}{*}{$E_{0,\text{s}}^{(t)}(s)$} & $ [0,\infty] $ & $<0$ & & & & & \\
			
			\midrule	
			
			& $[-1,0]$ & $\geq0$ & $\times$ & \multirow{2}{*}{$\cap$} & \multirow{2}{*}{$\checkmark$}  & \multirow{2}{*}{$-H(X|B)_\rho$} & \multirow{2}{*}{$V^{(t)}(P,\mathscr{W})$} \\
			\multirow{-2}{*}{$E_{0,\text{s}}^{(t)}(s,P)$} & $  [0,\infty] $ & $<0$ & $\searrow$ &  &  & & \\		
			
			\bottomrule				
		\end{tabular}
	}
	
\caption{
The table compares properties of different types of the auxiliary functions in finite-dimensional Hilbert space. The functions $E_{0}^{\texttt{r},(t)}(s,P)$, $E_{0}^{\texttt{a},(t)}(s,P)$, $E_{0,\text{s}}^{(t)}(s)$, and $E_{0,\text{s}}^{(t)}(s,P)$, respectively, correspond to the auxiliary functions of R\'enyi information, Augusting information in classical-quantum channel coding, and the auxiliary functions in classical data compression with quantum side information of i.i.d.~sources \cite{CHDH-2018}, and type-dependent sources \cite{CHDH2-2018}. The three values of $(t) = \{\,\}$, $*$, and $\flat$ denote the Petz's, sandwiched, and log-Euclidean R\'enyi divergence. In the last two columns, we assume the auxiliary functions are second-order differentiable with respect to $s$.
	The information variance quantities are defined by $V^{(t)}(P,\mathscr{W}) = \sum_{x\in\mathcal{X}} P(x) V^{(t)}(W_x\|P\mathscr{W})$, $
	U^{(t)}(P,\mathscr{W}) = V^{(t)}(P\circ \mathscr{W}\| P\otimes P\mathscr{W})$, and $
	V^{(t)}(X|B)_\rho = V^{(t)}( \rho_{XB} \| \mathds{1}_X\otimes \rho_B)$.
	We refer the readers to Sections~\ref{sec:preliminaries} and \ref{sec:properties} for detailed definitions.
	}	\label{table:comparison}	
\end{table}

\section{Preliminaries and Notation} \label{sec:preliminaries}

Throughout this paper, we consider a complex separable Hilbert space $\mathcal{H}$, and let $|\mathcal{H}|$ denote its dimension.
Let $\mathcal{B(H)}$ and $\mathcal{L}_{> 0}(\mathcal{H})$ denote the algebra of bounded linear operators and {non-zero} positive semi-definite operators on $\mathcal{H}$. 
For $0<p\le \infty $, we denote by $\|M\|_p := \left( \Tr | M |^p \right)^{\sfrac{1}{p} }$ the Schatten $p$-norm for $\mathcal{B}(\mathcal{H})$ and also the $L_p$-norm for a von Neumann algebra $(\mathcal{M},\text{Tr})$.
The Schatten $p$ class on $\mathcal{H}$ is denoted by $S_p(\mathcal{H}):= \{ M \in \mathcal{B(H)}: \|M\|_p <\infty  \}$.
We use $\mathcal{S(H)}$  to denote the set of density operators (i.e.~positive semi-definite operators with unit trace) on $\mathcal{H}$. 
We also use $\mathcal{S(M)}$ to denote the density operators of a von
Neumann algebra $(\mathcal{M},\Tr)$ equipped with the trace $\Tr$. Namely, $\mathcal{S(M)}$ is a subset of $L_1$-space $L_1(\mathcal{M})$ consisting of all positive and unit trace operators.
Here we have slightly abused the notation that $\mathcal{S(M)}$ is the (normal) state space of von Neumann algebra $\mathcal{M}$, while the notation $\mathcal{S(H)}:= \mathcal{S(B(H))}$ means the (normal) state space of $\mathcal{B(\mathcal{H})}$, which is a subset of the trace class operator $S_1(\mathcal{H})$ but not a subset of the Hilbert space $\mathcal{H}$.
The symbol $\mathds{1}_\mathcal{H}$ denotes the identity operator in $\mathcal{B(H)}$.
We denote by $\mathscr{P}(\mathcal{X})$ the set of all probability measures on a finite set $\mathcal{X}$.
 For two real-valued functions $f$ and $g$, $f\vee g $ is the pointwise maximum of $f$ and $g$, and $f\wedge g$ is the pointwise minimum. We use $\textsf{supp}(A)$ to denote the support of an operator or a function $A$. We use $\mathrm{i}$ to denote the imaginary unit.

\subsection{Quantum Entropies} \label{sec:entropy}

For $\rho,\sigma\in\mathcal{L}_{> 0}(\mathcal{H})$ and $\alpha\in (0,\infty)\backslash\{1\}$, the Petz's R\'enyi divergence~\cite{Pet86}, sandwiched R\'enyi divergence~\cite{MDS+13,WWY14, BST18, Jen18a}, and log-Euclidean R\'enyi divergence~\cite{ON00, MO17} are defined as\footnote{The three quantities are finite when $\rho\ll \sigma$, or $\rho$ is not orthogonal to $\sigma$ and $\alpha<1$. Otherwise, they are positive infinite.}
\begin{align}
&D_\alpha(\rho\|\sigma) := \frac{1}{\alpha-1}\log \frac{\Tr[\rho^\alpha\sigma^{1-\alpha}]}{\Tr\left[\rho\right]},\\
&D_\alpha^*(\rho\|\sigma) := \frac{1}{\alpha-1}\log \frac{ \Tr\left[ \left(\sigma^{\frac{1-\alpha}{2\alpha}} \rho \sigma^{\frac{1-\alpha}{2\alpha}} \right)^\alpha \right] }{\Tr\left[\rho\right]}, \label{eq:D*} \\
&D^\flat_\alpha(\rho\|\sigma) := \frac{1}{\alpha-1} \log \frac{\Tr \left[\mathrm{e}^{\alpha \log \rho + (1-\alpha) \log \sigma}\right]}{{\Tr\left[\rho\right]}}.
\end{align}
It is known \cite{Mos15, MDS+13, WWY14} that all the $\alpha$-R\'enyi divergences converge to the Umegaki relative entropy \cite{Ume62} 
\begin{align*}
D(\rho\|\sigma) := \frac{1}{\Tr[\rho]}\Tr\left[ \rho(\log \rho - \log \sigma)\right]
\end{align*}
as $\alpha\to 1$, i.e.~
\begin{align}
D^{(t)}_1(\rho\|\sigma) := \lim_{\alpha\to 1} D_\alpha^{(t)}(\rho\|\sigma) = D(\rho\|\sigma).
\end{align}
The cases of $D_0^{(t)}$ and $D_\infty^{(t)}$ are defined as limits of $D_\alpha^*$ for $\alpha\to\{0,\infty \}$.
The following Lemma~\ref{lemm:prop} collects useful properties of the $\alpha$-R\'enyi divergence and ensures the existence of $D_0^{(t)}$ and $D_\infty^{(t)}$.

\begin{lemm}[Properties of order $\alpha$ R\'enyi Divergences] \label{lemm:prop}
	The following holds.
	\begin{enumerate}[(a)]
		\item\label{prop-a}
		Let $(t) = \{\,\}, *$ or $\flat$.
		For any $\rho \in \mathcal{S(H)}$, the map $\alpha \to D_\alpha^{(t)}(\rho\|\sigma)$ is continuous and nondecreasing on $[0,\infty]$.
		
		\item\label{prop-b}
		Let $(t) = \{\,\}, *$ or $\flat$.
		For $\rho,\sigma\in\mathcal{L}_{> 0}(\mathcal{H})$, and $\alpha \in (0,\infty]$, we have $D_\alpha^{(t)}(\rho\|\sigma) \geq \log \Tr\left[\rho\right]-\log \Tr[\sigma]$ with equality if and only if $\rho$ is a constant multiple of $\sigma$.
		Moreover, $D_\alpha^{(t)}(\rho\|\sigma) \geq 0$ for $\rho,\sigma \in \mathcal{S(H)}$ with equality if and only if $\rho = \sigma$.

		\item\label{prop-c} For any $\rho,\sigma_1,\sigma_2\in\mathcal{L}_{> 0}(\mathcal{H})$ with $\sigma_1\leq \sigma_2$, we have $ D_\alpha^{(t)}(\rho\|\sigma_1)\geq D_\alpha^{(t)}(\rho\|\sigma_2)$ for $(t) = \{\,\}$ and $\alpha\in[0,1]$, for $(t)=*$ and $\alpha\in[1/2, \infty]$, and for $(t) = \flat$ and $\alpha\in[0,\infty]$.
		
		\item\label{prop-d} For any $\rho \in  \mathcal{L}_{> 0}(H)$, $ D_\alpha^{(t)}(\rho\|\sigma)$ is convex on $\mathcal{S(H)}$ for $(t) = \{\,\}$ and $\alpha\in[0,2]$, for $(t) = *$ and $\alpha\in[1/2, \infty]$.
		and for $(t) = \flat$ and $\alpha\in[0,\infty]$
		
		\item\label{prop-e} Let $(t) = \{\,\}, *$ or $\flat$. For any $\alpha\in (0,\infty]$ and $\rho \in \mathcal{L}_{> 0}(\mathcal{H})$, $\sigma\mapsto D_\alpha^{(t)}(\rho\|\sigma)$ is lower semi-continuous on $\mathcal{S(H)}$.
		
		\item \label{prop-f} For any $\rho , \sigma \in  \mathcal{L}_{> 0}(H)$, we have
		\begin{align}
		&D_\alpha^*(\rho\|\sigma) \leq D_\alpha(\rho\|\sigma) \leq D_\alpha^\flat (\rho\|\sigma), \quad \alpha\in [0,1], \\
		&D_\alpha^\flat(\rho\|\sigma) \leq D_\alpha^*(\rho\|\sigma) \leq D_\alpha (\rho\|\sigma), \quad \alpha\in [1,\infty]. \label{eq:temp}
		\end{align}
	\end{enumerate}
\end{lemm}

We note that
\ref{prop-a} was proved in \cite[Lemma~3.12, Corollary~3.15]{MO17} and \cite[Theorem 7]{MDS+13};
\ref{prop-b} was shown in \cite[Theorem 3]{MDS+13}, \cite[Theorem 5]{Bei13} and \cite[Proposition~3.22]{MO17};
\ref{prop-c} was proved in \cite[Proposition~4]{MDS+13} and \cite[Lemma~3.24]{MO17};
\ref{prop-d} was shown in \cite[Proposition~3.18]{MO17}\footnote{
We note that in Ref.~\cite{MDS+13} the definitions for the three types of R\'enyi divergences do not have the scaling factor $\Tr[\rho]$ in the denominator.
Here, we adopt the notation introduced in Ref.~\cite{MO17}.
For \cite[Proposition~3.18]{MO17} (corresponding to Lemma~\ref{lemm:prop}-\ref{prop-d}) and \cite[Corollary~3.27]{MO17} (corresponding to Lemma~\ref{lemm:prop}-\ref{prop-e}), we remark that the proofs work for both definitions of the R\'enyi divergences.
},
\ref{prop-e} was proved in \cite[Theorem~15]{EH14}, \cite[Corollary~3.27]{MO17};
and \ref{prop-f} is a consequence of the Araki-Lieb-Thirring inequality \cite{LT76, Ara90} and Golden-Thompson inequality \cite{Gol65, Sym65, Tho65} (see also \cite[Proposition~3.20]{MO17}).

Finally, for $\rho,\sigma \in \mathcal{S(H)}$, the {quantum relative entropy variances} \cite{TH13, Li14} are defined as
\begin{align}
&V^{(t)}(\rho\|\sigma) :=  \Tr \left[ \rho \left( \log \rho - \log \sigma \right)^2 \right] - D(\rho\|\sigma)^2, \quad (t)=\{\,\} \text{ or } * \\
&{V}^\flat(\rho\|\sigma) :=
\int_{0}^1\mathrm{d} t\Tr\left[ \rho^{1-t} (\log \rho - \log \sigma ) \rho^t (\log \rho - \log \sigma ) \right] - D(\rho\|\sigma)^2. \label{eq:relative_V}
\end{align}



\section{Complex Interpolation and Noncommutative {$L_p$} Spaces} \label{sec:interpolation}

In this section, we first recall the definition of the complex interpolation and the noncommutative $L_p$ spaces. The main result in this section is an interpolation inequality corresponding to a quasi-norm defined in Eq.~\eqref{eq:Petz_norm}.
This will be employed to prove the concavity of Petz's auxiliary function in Section~\ref{sec:properties}.

Let us start with complex interpolation. We refer to \cite{BL76} for a detailed account of interpolation spaces.  Let $X_0$ and $X_1$ be two Banach spaces. Assume that there exists a Hausdorff topological vector space $X$ such that $X_0, X_1\subset X$ as subspaces. Let $\mathcal{Z}=\{z\,|\,0\le \textsf{Re} (z)\le 1\}$ be the unit vertical strip on the complex plane, and $\mathcal{Z}_0=\{z\,|0< \textsf{Re} (z)< 1\}$ be its open interior. Let $\F(X_0, X_1)$ be the space of all functions $f:\mathcal{Z}\to X_0+X_1$, which are bounded and continuous on $\mathcal{Z}$ and analytic on $\mathcal{Z}_0$, and moreover
\[\{f(\mathrm{i}t)\pl|\pl t\in \mathbb{R}\}\subset X_0\pl ,\pl \{f(1+\mathrm{i}t)\pl|\pl t\in \mathbb{R}\}\subset X_1\pl.\]
$\F(X_0, X_1)$ is again a Banach space equipped with the norm
\[\norm{f}{\F} :=\max\left\{\pl \sup_{t\in \mathbb{R}} \norm{f(\mathrm{i}t)}{X_0}\pl,\pl \sup_{t\in \mathbb{R}}\norm{f(1+\mathrm{i}t)}{X_1}\right\}\pl. \]
The complex interpolation space $(X_0,X_1)_\theta$, for $0\le \theta\le 1$, is the quotient space of $\F(X_0,X_1)$ as follows,
\[(X_0, X_1)_\theta=\{\pl x\in X_0+X_1\pl| \pl x=f(\theta) \pl \text{for some }\pl  f\in \F(X_0, X_1)\pl\} \pl.\]
where quotient norm is
\begin{align}\label{def}\norm{x}{\theta}=\inf \{\pl\norm{f}{\F}\pl| \pl f(\theta)=x \pl\}\pl .\end{align}
It is clear from the definition that $X_0=(X_0,X_1)_0,  X_1=(X_0,X_1)_1$. For all $0<\theta<1$,
$(X_0,X_1)_\theta$ are called interpolation space of $(X_0,X_1)$. A consequence of the definition \eqref{def} is the \emph{interpolation inequality} that for an analytic function $f:\mathcal{Z}\to X_0+X_1$,
\begin{align} \norm{f(\theta)}{(X_0,X_1)_\theta}\le \left(\sup_{t\in\mathbb{R}}\norm{f(\mathrm{i}t)}{X_0}\right)^{1-\theta}\left(\sup_{t\in\mathbb{R}}\norm{f(1+\mathrm{i}t)}{X_1}\right)^{\theta}\pl. \label{in}\end{align}
This follows from applying the definition to the analytic function \[\tilde{f}(z)=\left(\sup_{t\in\mathbb{R}}\norm{f(\mathrm{i}t)}{X_0}\right)^{z-1}\left(\sup_{t\in\mathbb{R}}\norm{f(1+\mathrm{i}t)}{X_1}\right)^{-z}f(z).\]

The most basic example is that the $p$-integrable function spaces $L_p(\Omega,\mu)$ of a positive measure space $(\Omega,\mu)$. $L_p(\Omega,\mu)$ for $1\le p\le \infty$ forms a family of interpolation spaces, i.e.
\[ L_p(\Omega,\mu)\cong [L_{p_0}(\Omega,\mu),{L_{p_1}}(\Omega,\mu)]_\theta\]
holds isometrically for all $1\le p_0,p_1,p\le \infty, 0\le \theta\le 1$ such that
$\frac{1}{p}=\frac{1-\theta}{p_0}+\frac{\theta}{p_1}$. For a von Neumann algebra $(\mathcal{M},\text{Tr})$ equipped with normal faithful semifinite trace $\text{Tr}$,
the noncommutative $L_p$-norm is defined as $\|x\|_p=\text{Tr}(|x|^p)^{\frac{1}{p}}$ and $L_p(\mathcal{M},\text{Tr})$ (or shortly $L_p(\mathcal{M})$) is the completion of $\{x\in \iM \pl |\pl \|x\|_p<\infty\}$. The analog of \eqref{in} is that
\[ L_p(\mathcal{M},\text{Tr})\cong [L_{p_0}(\mathcal{M},\text{Tr}),{L_{p_1}}(\mathcal{M},\text{Tr})]_\theta .\]
In particular, the Schatten-$p$ class on a Hilbert space $\mathcal{H}$ satisfies
\[S_p(\mathcal{H}) \cong \left[S_{p_0}(\mathcal{H}), S_{p_1}(\mathcal{H}) \right]_{\theta}\pl.\]
Here $S_\infty(\mathcal{H}):= \mathcal{B(H)}$ denotes the bounded operators on $\mathcal{H}$. The interpolation relation has already been used in many works in quantum information theory, e.g.~\cite{WWY14}.

In \cite{JP10} Junge and Parcet introduced the amalgamated $L_p$-space for a subalgebra $\iN\subset \iM$. Here, for simplicity, we consider the case $(\iM, \Tr)$ being a semi-finite von Neumann algebra and $\iN$ is a subalgebra of $\mathcal{M}$ such that $\Tr_{|\iN}$ is also semi-finite.
Let $1\le p\le\infty $ and $\frac{1}{p}+\frac{1}{p'}=1$. For $x\in \iM$, the amalgamated $L_{1}^p$ norm of the inclusion $\iN\subset \iM$ is defined as
\begin{align}\norm{x}{L_{1}^p(\iN\subset\iM)} :=\inf\{\norm{a}{2p'} \norm{y}{p}\norm{b}{2p'}\pl |\pl x=a yb\}\pl ,
\label{augmented}\end{align}
where the infimum is taken over all factorizations $x=ayb$ such that $a,b\in L_{2p'}(\iN)$ and $y\in L_p(\iM)$. When $x$ is positive, the above expression simplifies to
\begin{align}\norm{x}{L_{1}^{p}(\iN\subset\iM)} =\inf_{\si \in \iS(\iN)} \left\| \si^{-\frac{1}{2p'}}x \si^{-\frac{1}{2p'}} \right\|_p.
\label{1augmented}\end{align}
Here and in the following, the infimum is taken over all density operators $\si \in \mathcal{N}$ such that $x=\si^{\frac{1}{2p'}}y\si^{\frac{1}{2p'}}$ with $y\in L_p(\M)$ so that $\si^{-\frac{1}{2p'}}x \si^{-\frac{1}{2p'}}$ is well defined. For example, let $\iN=\mathds{1}_{\mathcal{H}_A} \ten \mathcal{B}(\mathcal{H}_B)$ and $\iM= \mathcal{B}(\mathcal{H}_A) \otimes \mathcal{B}(\mathcal{H}_B)$ equipped with the usual matrix trace on $\mathcal{H}_A$ and $\mathcal{H}_A\ten \mathcal{H}_B$, respectively. This is the $L_p$-norm corresponding to sandwiched conditional R\'enyi entropy \cite{MDS+13,WWY14, Tom16}:
\begin{align} \label{eq:conditional}
\begin{split}
{H}_p^*(A|B) &:= -  \inf_{\sigma_B \in \mathcal{B}(\mathcal{H}_B) } D_p^*(\rho_{AB} \| \mathds{1}_{\mathcal{H}_A} \otimes \sigma_B) \\
&= -p'\log \norm{\rho^{AB}}{L_1^p(\mathds{1}_{\mathcal{H}_A} \ten \mathcal{B}(\mathcal{H}_A) \subset \mathcal{B}(\mathcal{H}_A) \otimes \mathcal{B}(\mathcal{H}_B))}\pl.
\end{split}
\end{align}
This special case was introduced and studied by Pisier in \cite{Pis98} as vector-valued noncommutative $L_p$-spaces. In general, for an inclusion $\iN\subset \iM$, Junge and Parcet proved that
\begin{enumerate}
	\item[i)] $\norm{\cdot}{L_1^p(\iN\subset \iM)}$ is indeed a Banach space norm for $1\le p\le \infty$.
	\item[ii)] for all $1\le p_0,p_1,p\le \infty, 0\le \theta\le 1$  such that $\displaystyle \frac{1}{p}=\frac{1-\theta}{p_0}+\frac{\theta}{p_1}$,
	\begin{align} \label{eq:inter_sandwiched}
	L_{1}^{p_0}(\iN\subset\iM),L_{1}^{p_1}(\iN\subset\iM)]_\theta \cong L_{1}^{p}(\iN\subset\iM) \pl
	\end{align}
	holds isometrically.
\end{enumerate}
We also consider the expression corresponding to Petz's version. We define that for a positive $x$,
\begin{align}
\left\|x\right\|_{ \bar{L}_1^p(\mathcal{N}\subset\mathcal{M})  } := \inf_{\sigma\in \iS(\mathcal{N})} \left(\Tr\left[  x^p \sigma^{1-p} \right] \right)^{\frac1p}. \label{eq:Petz_norm}
\end{align}
 We have the following Sibson identity \cite{Sib69, SW12} for a subalgebra $\iN\subset \iM$. Recall that the conditional expectation $\mathbb{E}:\mathcal{M} \to \mathcal{N}$ is a unique completely positive trace-preserving map such that
\[\Tr(xa)=\Tr(\mathbb{E}(x)a)\ , \mathbb{E}(axb)=a \mathbb{E}(x)b\ , \ \forall a,b\in \iN, x\in \iM \ .\]

\begin{prop}
	[Noncommutative Sibson Identity] \label{prop:Sibson}
	Let $\iM$ be a finite von Neumann algebra and $\mathcal{N}\subset \mathcal{M}$ be a subalgebra. For all $\rho\in \mathcal{S(M)},\sigma\in \mathcal{S(N)}$, and $\alpha \in (0,\infty)$, it follows that
	\begin{align}
	D_\alpha\left(\rho\|\sigma \right) &= D_\alpha(\sigma^\star \| \sigma) + \frac{\alpha}{\alpha-1}\log \Tr\left[ \left( \mathbb{E}(\rho^\alpha) \right)^{\frac{1}{\alpha}} \right] \\
	&= D_\alpha(\sigma^\star \| \sigma) + 	 D_\alpha\left(\rho\|\sigma^\star \right),
	\end{align}
	where
	\begin{align}
	\sigma^\star := \frac{  \left( \mathbb{E}(\rho^\alpha)\right)^{\frac{1}{\alpha}} }{ \Tr\left[ \left( \mathbb{E}(\rho^\alpha) \right)^{\frac{1}{\alpha}} \right] }.
	\end{align}
	In particular, $\displaystyle \inf_{\sigma\in \mathcal{S(N)}} D_\alpha\left(\rho\|\sigma \right)= \frac{\alpha}{\alpha-1}\log \Tr\left[ \left( \mathbb{E}(\rho^\alpha) \right)^{\frac{1}{\alpha}} \right]$
\end{prop}
\begin{proof} Using the property of conditional expectation,
	\begin{align}
	D_\alpha\left(\rho\|\sigma \right) &= \frac{1}{\alpha-1} \log \Tr\left[ \rho^\alpha \sigma^{1-\alpha} \right] \\
	&= \frac{1}{\alpha-1} \log \Tr\left[ \mathbb{E}(\rho^\alpha) \sigma^{1-\alpha} \right] \\
	&= \frac{1}{\alpha-1} \log \Tr\left[ (\sigma^\star)^\alpha \sigma^{1-\alpha} \right] + \frac{\alpha}{\alpha-1}\log \Tr( \mathbb{E}(\rho^\alpha)^{\frac{1}{\alpha}}) \\
	&= D_\alpha(\sigma^\star \| \sigma) + \frac{\alpha}{\alpha-1}\log \Tr( \mathbb{E}(\rho^\alpha)^{\frac{1}{\alpha}}).
	\end{align}
	Note that
	\[\Tr( \mathbb{E}(\rho^\alpha)^{\frac{1}{\alpha}})=\Tr( \mathbb{E}(\rho^\alpha) \mathbb{E}(\rho^\alpha)^{\frac{1-\alpha}{\alpha}})=\Tr( \rho^\alpha \mathbb{E}(\rho^\alpha)^{\frac{1-\alpha}{\alpha}})=\Tr( \rho^\alpha (\si^\star)^{1-\alpha})\Tr( \mathbb{E}(\rho^\alpha)^{\frac{1}{\alpha}})^{1-\alpha}\ .\]
	Thus
	\[\frac{\alpha}{\alpha-1}\log \Tr( \mathbb{E}(\rho^\alpha)^{\frac{1}{\alpha}})=D_\al(\rho||\sigma^\star).\]
	The last assertion follows form the non-negativity of Petz's R\'enyi divergence $D_\alpha(\sigma^\star \| \sigma)$ (see e.g.~\cite{BST18}).
\end{proof}
\begin{remark}Proposition~\ref{prop:Sibson} is a generalization of the quantum Sibson identity proved by Sharma and Warsi \cite{SW12} for the case
	$\mathcal{M} = \mathcal{B}(\mathcal{H}_A\ten \mathcal{H}_B)$, $\mathcal{N} = \mathds{1}_A \otimes \mathcal{B}(\mathcal{H}_B)$ and $\mathbb{E}$ is the partial trace on system $A$.
	{We observe that the quantum Sibson identity can be interpreted from a more general framework of noncommutative measure space with conditional expectation.}	
\end{remark}

By using Proposition~\ref{prop:Sibson}, we can rewrite for all $p\geq 1$,
\begin{align}
\left\|x \right\|_{ \bar{L}_1^p(\mathcal{N}\subset\mathcal{M})  } =
\Tr\left[\mathbb{E}(x^p)^{\frac{1}{p}} \right]. \label{eq:E_Petz}
\end{align}
For this quantity, we have the following interpolation type inequality. This might be of  independent interest.
\begin{prop}
	[The interpolation inequality for $\left\| \,\cdot\, \right\|_{ \bar{L}_1^p(\mathcal{N}\subset\mathcal{M})  }$] \label{theo:inter_Petz}
	For every $x\in\mathcal{M}$, $1< p_0, p_1, p< \infty$, $0\leq \theta\leq 1$ such that $\frac1p = \frac{1-\theta}{p_0} + \frac{\theta}{p_1}$, it holds that
	\begin{align} \left\|{x}\right\|_{\bar{L}_{1}^{p}(\mathcal{N} \subset\mathcal{M}) }\le\left\|{x}\right\|_{\bar{L}_{1}^{p_0}(\mathcal{N}\subset\mathcal{M}) }^{1-\theta}\left\|{x}\right\|_{\bar{L}_{1}^{p_1}(\mathcal{N}\subset\mathcal{M}) }^\theta \pl. \label{eq:inter_Petz}
	\end{align}
\end{prop}
\begin{proof}
[Proof of Proposition~\ref{theo:inter_Petz}]
	
Denote $\gamma = \frac{p(1-\theta)}{p_0}$ and $1-\gamma = \frac{p\theta}{p_1}$.
	Let us first consider the case $\gamma = \frac12$ and $p = \frac12(p_0 + p_1)$.
	In this case, the inequality, Eq.~\eqref{eq:inter_Petz}, is equivalent to that for any $\sigma_0, \sigma_1 \in S(\mathcal{N})$, 
	\begin{align}
	\left\| x^{\frac{p_0}{2}} \sigma_0^{\frac{1-p_0}{2}} \right\|_2
	\left\| x^{\frac{p_1}{2}} \sigma_1^{\frac{1-p_1}{2}} \right\|_2
	\geq 	\left\| \mathbb{E}(x^p) \right\|_{\frac1p}.
	\end{align}

	Starting with Cauchy-Schwartz inequality,
	\begin{align}
	\left\| x^{\frac{p_0}{2}} \sigma_0^{\frac{1-p_0}{2}} \right\|_2
	\left\| x^{\frac{p_1}{2}} \sigma_1^{\frac{1-p_1}{2}} \right\|_2
	&\geq \left\| \sigma_1^{\frac{1-p_1}{2}} x^{\frac{p_1}{2}} x^{\frac{p_0}{2}} \sigma_0^{\frac{1-p_0}{2}} \right\|_1 \\
	&= \left\| \sigma_1^{\frac{1-p_1}{2}} x^p \sigma_0^{\frac{1-p_0}{2}} \right\|_1 \\
	&\geq \left\| \mathbb{E}\left(  \sigma_1^{\frac{1-p_1}{2}} x^p   \sigma_0^{\frac{1-p_0}{2}}  \right) \right\|_1 \label{eq:ip1} \\
	&= \left\|  \sigma_1^{\frac{1-p_1}{2}} \mathbb{E}\left(  x^p   \right) \sigma_0^{\frac{1-p_0}{2}}  \right\|_1 \label{eq:ip2} \\
	&\geq \left\|  \mathbb{E}\left( x^p \right) \right\|_{\frac{1}{p}}, \label{eq:ip3}
	\end{align}
	where inequality ~\eqref{eq:ip1} follows from the fact that the conditional expectation $\mathbb{E}$ is a contraction for $1$-norm, and Eq.~\eqref{eq:ip2} uses the module property of $\mathbb{E}$, i.e.~
	\begin{align}
	\mathbb{E}\left(axb \right) = a \mathbb{E}(x) b, \quad \forall a,b\in\mathcal{N},\; x \in \mathcal{M}.
	\end{align}
	The last inequality \eqref{eq:ip3} is the H{\"{o}}lder inequality for $p = 1 + \frac{p_0-1}{2} + \frac{p_1-1}{2}$,
	\begin{align}
	\left\|  \mathbb{E}\left( x^p \right) \right\|_{\frac{1}{p}}
	&\leq \left\| \sigma_1^{\frac{p_1-1}{2}} \right\|_{\frac{2}{p_1-1}} \left\| \sigma_1^{\frac{1-p_1}{2} } \mathbb{E}(x^p) \sigma_0^{\frac{1-p_0}{2}} \right\|_1 \left\| \sigma_0^{\frac{p_0-1}{2}} \right\|_{\frac{2}{p_0-1}} \\
	&\leq \left\| \sigma_1^{\frac{1-p_1}{2} } \mathbb{E}(x^p) \sigma_0^{\frac{1-p_0}{2}} \right\|_1.
	\end{align}
	Here, $\left\|\sigma_1\right\|_1 = \left\|\sigma_0\right\|_1 = 1$ because they are density operators.
	
	This proves the inequality for $\gamma = \frac12$. Using induction, we obtain the inequality for $2^n$-partition points $p = k 2^{-n}(p_1-p_0) + p_0$ for all $k,n\in\mathbb{N}$, $0\leq k \leq 2^n$. The case for general $p$ follows from the continuity.
\end{proof}

\begin{remark}
	For finite-dimensional matrices, the above interpolation inequality, Eq.~\eqref{eq:inter_Petz}, is a special case of \cite[Corollary 3.7]{BL16} with trace norm: for all unitary-invariant norms $\vertiii{\,\cdot\,}$ and $\nu >0$, the map
		\begin{align}
		(p,t) \mapsto \vertiii{  \left( \Lambda(A^{\frac{t}{p}}) \right)^{\nu p} }
		\end{align}
		is jointly log-convex on $(0,\infty)\times (-\infty, \infty)$ for any positive linear maps $\Lambda$ on positive semi-definite matrices. We remark that for von Neumann algebras similar results have been studied by Shao \cite{Sha17}. In particular, our inequality for conditional expectation can also be derived from \cite[Corollary 3.13]{Sha17} via an averaging trick.
\end{remark}

\section{Noncommutative R\'enyi and Augustin Information} \label{sec:information}
In this section, we introduce the R\'enyi information and Augustin information. These quantities are usually defined for a channel. However, in the case of classical-quantum channels, the channel output can be viewed as a collection of density operators or noncommutative measures. We  can thus define the R\'enyi and Augustin information from a perspective of noncommutative measure space without introducing the classical-quantum channels. The connection can be easily understood, and we delay this until Section~\ref{sec:applications}.

We firstly establish fundamental properties for both the R\'enyi and Augustin information (Propositions~\ref{prop:I2_R} and \ref{prop:I2}). Secondly, in Section~\ref{sec:infimum} we study whether the infimum in the definition of the R\'enyi information can be attained.

Let $\mathscr{W} \subset \mathcal{S(H)}$ be a set of density operators on $\mathcal{H}$.
Given a prior probability mass function $P\in\mathscr{P(W)}$, and $\alpha\in[0,\infty]$, we define the \emph{order $\alpha$ R\'enyi  information} and the \emph{order $\alpha$ Augustin information}, respectively, by
\begin{align}
I_\alpha^{\texttt{r},(t)}(P,\mathscr{W}) &:= \inf_{\sigma\in\mathcal{S(H)}}  D_{\alpha}^{(t)}( P \circ \mathscr{W} \| P\otimes \sigma), \label{eq:I2_R} \\
I_\alpha^{\texttt{a},(t)}(P,\mathscr{W}) &:=
\inf_{\sigma\in\mathcal{S(H)}}  D_{\alpha}^{(t)}(\omega\|\sigma|P) :=
\inf_{\sigma\in\mathcal{S(H)}} \sum_{\omega \in\mathscr{W}} P(\omega) D_{\alpha}^{(t)}(\omega\|\sigma). \label{eq:I2}
\end{align}
for $(t) = \{\,\}$, $\{*\}$ and $\{\flat\}$. Here, $P\circ \mathscr{W} := \sum_{\omega} P(\omega) |\omega\rangle\langle\omega| \otimes \omega$ is a joint probability measure whose marginal distribution on the support of $P$ is $P$ and whose conditional distribution is $\omega \in \mathscr{W}$.
Here, we use the superscript `\texttt{r}' to indicate the R\'enyi information, while `\texttt{a}' to indicate the Augustin information.
As we defined for the R\'enyi divergence, the superscript `$(t)$' denotes that the information quantities are defined by the Petz $(t) = \{\}$, sandwiched $(t) = \{*\}$, or the log-Euclidean form $(t)= \{\flat\}$.

For $\alpha=1$, these two quantities correspond to the Holevo quantity \cite{Hol98}:
\begin{align}
I_1^{\texttt{r},(t)}(P,\mathscr{W}) = I_1^{\texttt{a},(t)}(P,\mathscr{W}) =
I(P,\mathscr{W}) :=
{D} \left( P\circ \mathscr{W} \| P\otimes P\mathscr{W} \right), \label{eq:mutual2}
\end{align}
where $P\mathscr{W} := \sum_{w\in \mathscr{W}} P(\omega) \omega$ is the marginal state on the output Hilbert space. If the measures are commutative, it is exactly Shannon's mutual information.

The order-$\alpha$ R\'enyi capacity is defined as follows \cite[Proposition 4.2, Corolary 4.5]{MO17}:
\begin{align}
\begin{split}\label{eq:radius}
C_{\alpha,\mathscr{W}}^{(t)} &:=
\sup_{P\in\mathscr{P(W)}} I_\alpha^{\texttt{r},(t)} (P,\mathscr{W})  \\
&= \sup_{P\in\mathscr{P(W)}} I_\alpha^{\texttt{a},(t)} (P,\mathscr{W})
\end{split}
\end{align}
for $(t)= \{\}$ and $\alpha\in[0,2]$, for $(t) = *$ and $\alpha \in [1/2,\infty]$, and for $(t) = \flat$ and $\alpha\in[0,\infty]$.


In the following Propositions~\ref{prop:I2_R} and \ref{prop:I2}, we prove important properties of the R\'enyi and Augustin information.

\begin{prop}[Properties of R\'enyi Information] \label{prop:I2_R}	
	Let $\mathscr{W} \subset \mathcal{S(H)}$, and let $(t)$ be any of the three values: $\{\}$, $*$, or $\flat$.
	\begin{enumerate}[(a)]
		\item\label{I2_R-a} For any $P\in\mathscr{P(W)}$, $I_\alpha^{\textnormal{\texttt{r}},(t)}(P,\mathscr{W})$ is non-negative and nondecreasing in $\alpha$. Moreover, $I_\alpha^{\textnormal{\texttt{r},}(t)}(P,\mathscr{W}) \leq\log |\textnormal{\textsf{supp}}(P)|$.
		
		\item\label{I2_R-b} The map $P\mapsto I_{\alpha}^{\textnormal{\texttt{r}},(t)} (P,\mathscr{W})$ is quasi-concave on $P\in\mathscr{P(W)}$ for $\alpha \in [0,1)$, and concave on $P\in\mathscr{P(W)}$ for $\alpha \in [1,\infty]$.
		
		\item\label{I2_R-c}	Let
		\begin{align} \label{eq:set_A}
		\mathcal{A} := [0,1],\quad \mathcal{A}^* := [1/2, \infty], \; \text{ and } \; \mathcal{A}^\flat := [0,\infty].
		\end{align}		
		For $\eta \in [0,\infty]$, if $C_{\eta, \mathscr{W}}^{(t)} < \infty$, then $\left\{ I_\alpha^{\textnormal{\texttt{r}},(t)}(P,\mathscr{W})  \right\}_{ \alpha\in[0,\eta] \cap \mathcal{A}^{(t)} }$ is uniformly equicontinuous in $P\in\mathscr{P(W)}$.
	\end{enumerate}
\end{prop}
\noindent The proof of Proposition~\ref{prop:I2_R} is given in Appendix~\ref{sec:proofs}.

\begin{prop}[Properties of Augustin Information] \label{prop:I2}
	Let $\mathscr{W}\subset \mathcal{S(H)}$ be a classical-quantum channel, $(t)$ be any of the three values: $\{\}$, $*$, or $\flat$, and let $\mathcal{A}^{(t)}$ be defined in \eqref{eq:set_A}.
	\begin{enumerate}[(a)]
		\item\label{I2-a} For every $P\in\mathscr{P}(\mathcal{X})$,
		$I_\alpha^{\textnormal{\texttt{a}},(t)}(P,\mathscr{W})$ is non-negative and nondecreasing in $\alpha$.
		Moreover, $I_\alpha^{\textnormal{\texttt{a}},(t)}(P,\mathscr{W}) \leq H(P)$ for $\alpha \in \mathcal{A}^{(t)}$, where $H(P)$ is the Shannon entropy of $P$.
		
		
		\item\label{I2-b} For any $\alpha> 0$, the map $P \mapsto I_{\alpha}^{\textnormal{\texttt{a}},(t)} (P,\mathscr{W})$ is concave on $P\in\mathscr{P(W)}$.
		
		\item\label{I2-c}
		For $\eta \in [0,\infty]$, if $C_{\eta, \mathscr{W}}^{(t)} < \infty$, then $\left\{ I_\alpha^{\textnormal{\texttt{a}},(t)}(P,\mathscr{W})  \right\}_{ \alpha\in (0,\eta] \cap \mathcal{A}^{(t)} }$ is uniformly equicontinuous in  $P\in\mathscr{P(W)}$.
		
	\end{enumerate}
	
\end{prop}

\noindent The proof of Proposition~\ref{prop:I2} is given in Appendix~\ref{sec:proofs}.

\subsection{Existence of Infimum} \label{sec:infimum}

The noncommutative Sibson identity established in Proposition~\ref{prop:Sibson} already guarantees that the infimum of $I_\alpha^{\texttt{r}}$ in Eq.~\eqref{eq:I2_R} is attained for all $\alpha\in(0,\infty)$.
The goal of this section is to show that the infimum of $I_\alpha^{\texttt{r},*}$ is attained for all $\alpha\in [1,\infty]$ (Corollary~\ref{coro:inf_sandwiched}).

We have already noted that the sandwiched R\'enyi information $I^{\texttt{r},*}_\al(P,\mathscr{W})$ is closely related to the noncommutative $L_1^\al$ space as follows,
\[I^{\texttt{r},*}_\al(P,\mathscr{W}) = \al'\log\norm{\rho}{S_1(\mathcal{H},l_\al^n)}\]
where $\rho=\oplus_{\omega}P(\omega)^{\frac{1}{\al}}\omega$, $l_\al^n$ is the $\al$-summable space for length $n$-sequence and in our case $\mathcal{H}$ is the separable infinite-dimensional Hilbert space. This is a special case of vector-valued $L_p$ space introduced by Pisier \cite{Pis98}. Recall that for a general element $x\in S_1(\mathcal{H},l_\al^n)$, the norm is defined as
\begin{align*}\norm{x}{S_1(\mathcal{H},l_\al^n)}=&\inf_{x=(a\ten \mathds{1}) y(b\ten \mathds{1})} \norm{a}{S_{2\al'}}\norm{y}{S_\al(\mathcal{H},l_\al^n)}\norm{b}{S_{2\al'}}\\=&\inf_{x=(a\ten \mathds{1}) y(b\ten \mathds{1})} \norm{a}{S_{2\al'}}\norm{y}{L_\al( \oplus_{i=1}^n \mathcal{B(H)})}\norm{b}{S_{2\al'}}\pl,\end{align*}
where the infimum takes all factorization $x=(a\ten \mathds{1}) y (b\ten \mathds{1})$ with $a,b\in S_{2\al'}$. For positive $x$, it suffices to consider $a=b\ge 0$, and then the norm can be rewritten as
\begin{align*}
\norm{x}{S_1(\mathcal{H}, l_\al^n)}=&\inf_{\si \in \mathcal{S(H)} }\norm{(\si^{-\frac{1}{2\al'}}\ten \mathds{1})x( \si^{-\frac{1}{2\al'}}\ten \mathds{1})}{\al},
\end{align*}
which links to the sandwiched R\'enyi information $I^{\texttt{r},*}_\al$.

For a general von Neumann algebra $\M$, the space $L_p(\M,l_\infty^n)$ has been studied in \cite{Junge02,JX07} for the purpose to understand the noncommutative martingale and maximal functions. In particular, \cite[Remark 3.7]{Junge02} states that for a positive $x$, the above infimum for $\al=\infty$ can be attained, which follows from a Grothendieck--Pietsch factorization theorem. Namely, for  $\rho=\oplus_\omega p(\omega)^{\frac{1}{\al}}\omega\in S_1( \mathcal{H} ,l_1^n)$, there exists a density $\sigma^\star \in \mathcal{S(H)}$ such that
\[
\norm{\rho}{S_1( \mathcal{H},l_\al^n)} = \inf_{\sigma\in \mathcal{S(H)} }
\left\| \left( \sigma^{-\frac{1}{2}}\ten \mathds{1} \right)\rho  \left( \sigma^{-\frac{1}{2}}\ten \mathds{1} \right) \right\|_\infty
= \sum_\omega P(\omega) \left\|  (\sigma^\star)^{-\frac{1}{2}} \omega  (\sigma^\star)^{-\frac{1}{2}} \right\|_\infty,
\]
which indicates the infimum in the definition of $I^{\texttt{r},*}_\infty$ is attained. Here, we show that such infimums are also attained for $1< \al< \infty$. Our argument uses the uniform convexity of noncommutative $L_p$ space.
Here let us start with recalling the duality relation that for $1\le p<\infty$ and $p'$ satisfying $\frac{1}{p}+\frac{1}{p'}=1$,
\[ L_p(\M)^*=L_{p'}(\M).\]
In particular, this implies that for $1<p<\infty$, $L_p(\M)^{**}=L_p(\M)$ are reflective. A stronger property for Banach spaces is the uniform convexity.
\begin{defn}A normed space $X$ is uniform convex if for any $0<\epsilon <2$, there exists a $\delta>0$ such that if $\norm{x}{}=\norm{y}{}=1$ and $\norm{\frac{x+y}{2}}{}\ge 1-\delta$, then $\norm{x-y}{}\le \epsilon.$
\end{defn}
For $1<p<\infty$, the uniform convexity of noncommutative $L_p(\M)$ follows from the Clarkson type inequality (see \cite{haagerup,fack,kosaki}, it is also clear that $L_1(\mathcal{M})$ and $L_{\infty}(\mathcal{M})$ are not uniform convex)
We will use the following consequence of uniform convexity.
\begin{lemm}\label{2} Let $1<p<\infty$ and $(x_n)_{n\in\mathbb{N}}$ be an infinite sequence of positive element in $L_{2p}(\iM)$. Suppose
	\begin{enumerate}[(a)]
		\item $\norm{x_n^2}{2p}=1$ for all $n\in\mathbb{N}$.
		\item $\displaystyle \lim_{N\to \infty}\inf_{n,m>N} \left\| \frac{x_n^2+x_{m}^2}{2} \right\|_p=1$.
	\end{enumerate}
	Then $(x_n)$ is a Cauchy sequence and hence $\displaystyle \lim_{n\to\infty} x_n=x \in L_{2p}(\iM)$ for some $x$ with $\norm{x}{2p}=1$.
\end{lemm}
\begin{proof}Given $\epsilon>0$, we choose $\delta>0$ such that if $\norm{a}{p}=\norm{b}{p}=1$
	and $\norm{\frac{a+b}{2}}{p}\ge 1-\delta$, then $\norm{a-b}{p} \le \epsilon$.
	By assumption, there exists $N$ large enough such that for all $n,m\ge N$,
	\[\left\|\frac{x_n^2+x_{m}^2}{2} \right\|_{p}\ge 1-\delta\pl,\]
	which implies
	\[\norm{x_n^2-x_m^2}{p}\le \epsilon\pl.\]
	Then the assertion follows from the inequality \cite[Lemma 2.1]{ricard}
	\[ \norm{x_n-y_m}{2p} \leq \norm{x_n^2-y_m^2}{p}^{\frac{1}{2}} \leq \epsilon.\]
\end{proof}
To show that the infimum in the definition of sandwiched R\'enyi information $I^{\texttt{r},*}_\al$ is attained, it is sufficient to consider that for any positive $x$, the factorization norm
\begin{align*}\norm{x}{S_1(l_\al)}=&\inf_{x=(a\ten 1) y(b\ten 1)} \norm{a}{S_{2\al'}}\norm{y}{\al}\norm{b}{S_{2\al'}}\\
=&\inf_{x=(a\ten \mathds{1}) y(a\ten \mathds{1})} \norm{a}{S_{2\al'}}^2\norm{y}{\al}
\\
=&\inf_{\sqrt{x}=(a\ten \mathds{1}) \eta} \norm{a}{S_{2\al'}}^2\norm{\eta}{2\al}^2
\end{align*}
is attained indeed by a factorization $x=(a\ten \mathds{1}) y (a\ten \mathds{1})=(a\ten \mathds{1}) \eta\eta^*(a\ten \mathds{1}) $.
\begin{prop}
	\label{p1}
	For $1< p<\infty$ and $x$ positive, the above infimum is attained, i.e. there exists $a\in S_{2p'}$ and $y\in S_p(l_p)$ such that $x=(a\ten \mathds{1}) y (a\ten 1)$ and
	\[\norm{x}{S_1(l_p)}=\norm{a}{2p'}^2\norm{y}{S_p(l_p)}.\]
\end{prop}
\begin{proof}Given $\norm{x}{S_1(l_p)}=1$, we have
	\[\inf_{\sqrt{x}=(a\ten 1) \eta} \norm{a}{S_{2p'}}\norm{\eta}{2p} =1\pl.\]
	We can find sequences $(a_n)\subset S_{2p'}$ and $(\eta_n)\subset S_p(l_p)$ such that for each $n$, $\sqrt{x}=(a_n\ten 1) \eta_n$, $\norm{a_n}{2p'}=1$ and \[\norm{\eta_n}{S_{2p}}\ge 1 \pl ,\pl \lim_{n\to \infty}\norm{\eta_n}{S_{2p}} = 1\pl.\]
	Without loss of generality, we can assume $a_n\ge 0$. Consider the $2\times 2$ factorization
	\[ \left[\begin{array}{cc}\sqrt{x} & 0 \\ 0&0\end{array}\right]=\left[\begin{array}{cc}\frac{a_n}{\sqrt{2}} & \frac{a_m}{\sqrt{2}}\\ 0&0\end{array}\right]\cdot\left[\begin{array}{cc}\frac{\eta_n}{\sqrt{2}} & 0 \\ \frac{\eta_m}{\sqrt{2}}&0\end{array}\right]\pl.\]
	Note that
	\begin{align*}
	&\left\| \left[\begin{array}{cc}\frac{a_n}{\sqrt{2}} & \frac{a_m}{\sqrt{2}}\\ 0&0\end{array}\right] \right\|_{2p'}= \left\|\frac{a_n^2+a_m^2}{2}\right\|_{p'}^{\frac{1}{2}}\\
	&\left\| \left[\begin{array}{cc}\frac{\eta_n}{\sqrt{2}} &0\\ \frac{\eta_m}{\sqrt{2}}&0\end{array}\right]\right\|_{2p}= \left\| \frac{\eta_n^*\eta_n+\eta_m^*\eta_m}{2} \right\|_{p}^{\frac{1}{2}}
	\end{align*}
	Because
	\[\left\| \left[\begin{array}{cc}x & 0 \\ 0&0\end{array}\right]\right\|_{S_1(l_p)}=\norm{x}{S_1(l_p)}=1,\]
	we know
	\[  \left\|\left[\begin{array}{cc}\frac{a_n}{\sqrt{2}} & \frac{a_m}{\sqrt{2}}\\ 0&0\end{array}\right]\right\|_{2p'} \left\| \left[\begin{array}{cc}\frac{\eta_n}{\sqrt{2}} &0\\ \frac{\eta_m}{\sqrt{2}}&0\end{array}\right]\right\|_{2p}\ge 1 .\]
	Moreover,  since
	\[ \left\| \frac{\eta_n^*\eta_n+\eta_m^*\eta_m}{2}\right\|_{p}^{\frac{1}{2}}\le \frac{1}{2}\norm{\eta_n}{2p}^2+\frac{1}{2}\norm{\eta_m}{2p}^2 \to 1.\]
	we have
	\[
	\lim_{N\to \infty} \sup_{n,m\ge N}\left\| \left[\begin{array}{cc}\frac{\eta_n}{\sqrt{2}} &0\\ \frac{\eta_m}{\sqrt{2}}&0\end{array}\right]\right|_{2p} \le 1,\]
	and hence \[\lim_{N\to \infty} \inf_{n,m\ge N}\norm{\frac{a_n^2+a_m^2}{2}}{p'} \ge 1\pl.\] By Lemma \ref{2}, we know $a_n\to a$ in $S_{2p'}$ with $\norm{a}{2p}=1$. On the other hand, because $S_{2p}(l_{2p})$ is reflexive, there exists a subsequence $\eta_{n_k}\to y$ weakly in the unit ball of $S_{2p}(l_{2p})$. Therefore, $\sqrt{x}=(a_{n_k}\ten 1)\eta_{n_k}\to (a\ten 1)\eta$ weakly and hence $\sqrt{x}=(a\ten 1)\eta$ and $\norm{\eta}{2p}=1$. That completes the proof.
\end{proof}

\begin{coro} \label{coro:inf_sandwiched}
	For each $\mathscr{W} \subset \mathcal{S(H)}$ and $P \in \mathscr{P}(\mathscr{W})$, the infimum for $I^{\textnormal{\texttt{r}},*}_\al(P,\mathscr{W})$ is uniquely attained for $1\le \al\le \infty$.
\end{coro}
\begin{proof}For $\al =1$, the infimum is attained by the (unnormalized) conditional expectation  onto $\mathcal{B(H)}$,
	\[\mathbb{E}(\oplus_{\omega} p(\omega)\omega )=\sum_{\omega}p(\omega)\omega\pl.\]
	For $\al =\infty$, it is discussed as above. For $1<\al<\infty$, it is equivalent to show that for some $\gamma\in S(\mathcal{H})$
	\begin{align*}
	\norm{\rho}{S_1(\mathcal{H},l_\al^n)}= &\norm{(\gamma^{-\frac{1}{2\al'}}\ten 1)\rho(\gamma^{-\frac{1}{2\al'}}\ten 1)}{\al}\\=&\norm{(\gamma^{-\frac{1}{2\al'}}\ten 1)\sqrt{\rho}}{2\al}^2,
	\end{align*} which follows from Proposition~\ref{p1}.
	The uniqueness is also guaranteed by the uniform convexity in Proposition~\ref{p1}.
\end{proof}

\begin{remark}
	Although we only focus on the sandwiched R\'enyi information in Corollary~\ref{coro:inf_sandwiched}, our approach given in Proposition~\ref{p1} applies to the sandwiched conditional R\'enyi entropy. Namely, the infimum in \eqref{eq:conditional} is uniquely attained for $1\leq p\leq \infty$.
\end{remark}

\section{Properties of Auxiliary Functions} \label{sec:properties}

This section is devoted to proving fundamental properties of the auxiliary functions, which are defined as a scaled version of the R\'enyi information and Augustin information introduced in Section~\ref{sec:information}.
Firstly, we present the convexity and concavity property of the auxiliary functions in prior probability distributions (Theorem~\ref{theo:prior}).
Secondly, we prove the concavity property in order $s\in(-1,0)$ for the sandwiched form (Theorem~\ref{theo:concave_*}) and for the Petz form (Theorem~\ref{theo:concave_Petz}), respectively, which thus answers the open problems as we described in Section~\ref{sec:introduction}.
Lastly, we employ the concavity and the equicontinuity of the R\'enyi and Augustin information established in Section~\ref{sec:information} to prove the joint continuity of the auxiliary functions (Theorem~\ref{theo:joint_cont}).
We note that results established in this section can be extended to tracial von Neumann algebras (see Remark~\ref{remark}). We present our proof in terms of bounded operator $\mathcal{B(H)}$ since the usual quantum information-theoretic protocols are formulated accordingly (see Section~\ref{sec:applications}).

Let $\mathscr{W} \subset \mathcal{S(H)}$ be an arbitrary set of density operators on $\mathcal{H}$.
Given $s>-1$, and a prior probability mass function $P\in\mathscr{P(W)}$, and $(t) \in \{\,\}, *$, and $\flat$,
we define the auxiliary functions for $\mathscr{W} $ in terms of the R\'enyi information and Augustin information:
\begin{align}
	E_{0}^{\texttt{r},(t)} (s,P) &:= s I_{\frac{1}{1+s}}^{\texttt{r},(t)}(P,\mathscr{W}) \label{eq:E0_c1} \\
	E_{0}^{\texttt{a},(t)} (s,P) &:= s I_{\frac{1}{1+s}}^{\texttt{a},(t)}(P,\mathscr{W}). \label{eq:E0_c2}
\end{align}
As in Section~\ref{sec:information}, we use superscript $\texttt{r}$ to denote the R\'enyi information, whereas we use $\texttt{a}$ to indicate the Augustin information.

\begin{theo}
	[Convexity/Concavity in Prior] \label{theo:prior}
	Let $\mathscr{W} \subset \mathcal{S(H)}$, and let $(t)$ be any of the three values.
	Then,
	\begin{enumerate}[(a)]
		\item $E_0^{\textnormal{\texttt{r}}, (t)}(s,P)$ is quasi-concave in $P$ for $s \geq0$, and convex in $P$ for $s\in[-1,0)$.
		
		\item $E_0^{\textnormal{\texttt{a}}, (t)}(s,P)$ is concave in $P$ for $s \geq0$, and convex in $P$ for $s\in[-1,0)$.
	\end{enumerate}
\end{theo}
\begin{proof}[Proof of Theorem~\ref{theo:prior}]
The assertions follow from item~\ref{I2_R-b} in Proposition~\ref{prop:I2_R}, item~\ref{I2-b} in Proposition~\ref{prop:I2}, and the definitions given in Eq.~\eqref{eq:E0_c1} and \eqref{eq:E0_c2}.
\end{proof}

In the following Theorem~\ref{theo:concave_*}, we establish the concavity of the sandwiched auxiliary functions $E_0^{\texttt{r},*}(s,P)$ and  $E_0^{\texttt{r},*}(s,P)$.

\begin{theo}[Concavity of the Sandwiched Form in Order] \label{theo:concave_*}
Let $\mathscr{W} \subset \mathcal{S(H)}$ and $P\in\mathscr{P(W)}$ be a probability mass function.
\begin{enumerate}[(a)]
\item\label{concave_*-a} 
The map $s\mapsto E_0^{\textnormal{\texttt{r}},*}(s,P)$ is concave on $(-1,0)$.
\item\label{concave_*b} If $\mathcal{H}$ is finite-dimensional, then $s\mapsto E_0^{\textnormal{\texttt{a}},*}(s,P)$ is concave on $(-1,0)$.
\end{enumerate}
\end{theo}

Before going into the details of the proof, we briefly sketch our proof strategy.
To show the concavity of $E_0^{\texttt{r},*}$, we introduce an analytic family of operators such that its amalgamated noncommutative norm coincides with the sandwiched R\'enyi information $I_\alpha^{\texttt{r},*}$. Then, the desired concavity follows from two facts: (i) the amalgamated noncommutative norm spaces form an interpolation family as we described in Section~\ref{sec:interpolation}, and (ii) the amalgamated noncommutative norms of the analytic family of operators are bounded at the boundary of the complex strip.

Regarding the $E_0^{\texttt{a},*}$, we are not allowed to apply the interpolation inequality with respect to amalgamated noncommutative norms because the set of tensor power of operators is not a subalgebra.
Instead, we directly prove an interpolation inequality for the target quantity by employing a technique of Devinatz’s factorization theorem~\cite{devinatz61}.

\begin{remark}
	In Ref.~\cite[Corollary 10]{TH13}, Tomamichel and Hayashi proved a concavity of the map $(\alpha - 1) \mapsto (1-\alpha) I_\alpha^{\texttt{r},*}(P,\mathscr{W})$, which is slightly different from the auxiliary function considered in Theorem~\ref{theo:concave_*}-~\ref{concave_*-a}, namely, $\alpha \mapsto E_0^{\texttt{r},*}(s,P)|_{s = 1/(1+\alpha)} = \frac{1-\alpha}{\alpha} I_\alpha^{\texttt{r},*}(P,\mathscr{W})  $.
	Their proof technique is the so-called \emph{pinching argument}, which allows the sandwiched R\'enyi information $I_\alpha^{\texttt{r},*}$ to inherit properties from the commuting case.
	Nevertheless, such the approach relies on the finite-dimensional assumption, whereas our method in Theorem~\ref{theo:concave_*}-\ref{concave_*-a} applies to infinite-dimensional Hilbert spaces as well as semi finite von Neumann algebras.
\end{remark}

\begin{proof}[Proof of Theorem~\ref{theo:concave_*}]	
	We first prove the concavity for $E_0^{\texttt{r},*}(s,P)$ and then $E_0^{\texttt{a},*}(s,P)$.
	Note that the sandwiched R\'enyi information can be written as:
\begin{align}
I_{\al}^{\texttt{r},*}(P,\mathscr{W})&= \inf_{\si\in\mathcal{S(H)}}  D^*_\al\left( \oplus_\omega P(\omega) \omega  || \oplus_\omega P(\omega)\si \right) \\
	&= \inf_{\si\in\mathcal{S(H)}} \frac{\alpha}{\alpha-1} \log \left\| \bigoplus_\omega \left( P(\omega) \si \right)^{\frac{1-\alpha}{2\alpha} } \left( P(\omega) \omega \right) \left( P(\omega) \si \right)^{\frac{1-\alpha}{2\alpha} } \right\|_{\alpha} \\
	&= \inf_{\si\in\mathcal{S(H)}} \frac{\alpha}{\alpha-1} \log \left\| \bigoplus_\omega P^{\frac{1}{\alpha}} (\omega) \si^{\frac{1-\alpha}{2\alpha} }  \omega \si^{\frac{1-\alpha}{2\alpha} } \right\|_{\alpha} \\
	&=\inf_{\si\in\mathcal{S(H)}} \frac{\alpha}{\alpha-1} \log \left( \sum_\omega P(\omega) \left\| \si^{\frac{1-\alpha}{2\alpha} }  \omega \si^{\frac{1-\alpha}{2\alpha}} \right\|_\alpha^\alpha \right)^{\frac{1}{\alpha}}.
\end{align}
Using the substitution $E_0^{\texttt{r},*}(s,P)|_{s = {(1-\alpha)}/{\alpha} } = \frac{1-\alpha}{\alpha} I_{\al}^{\texttt{r},*}(P,\mathscr{W})$, the concavity of $s\mapsto E_0^{\texttt{r},*}(s,P)$ for $s \in(-1,0)$, is equivalent to the log-convexity of the map:
\begin{align}
\frac{1}{\alpha} \mapsto  \inf_{\si\in\mathcal{S(H)}}\left( \sum_\omega P(\omega) \left\| \si^{\frac{1-\alpha}{2\alpha} }  \omega \si^{\frac{1-\alpha}{2\alpha}} \right\|_\alpha^\alpha \right)^{\frac{1}{\alpha}}, \quad \alpha > 1. \label{eq:log_convex_norm1}
\end{align}
To prove this, we	let $\mathcal{X} := \texttt{supp}(P)$,
$\mathcal{M} = \oplus_{x\in \mathcal{X}} \mathcal{B(H)}$ and
\begin{align}
\mathcal{N} = 1_\mathcal{X}\ten \mathcal{B(H)}:= \{a\oplus a \oplus \cdots \oplus a \ | \ a\in\mathcal{B(H)}  \} \subset \mathcal{M},
\end{align}
and let $\frac{1}{\alpha} = \frac{1-\theta}{\alpha_0} + \frac{\theta}{\alpha_1}$ for $\alpha_0,\alpha_1 \geq 1$ and $\theta \in [0,1]$.
Define the analytic family of operators,
\begin{align}
\rho(z) := \bigoplus_{\omega \in \mathscr{W}} P(\omega) ^{\frac{1-z}{\alpha_0}+ \frac{z}{\alpha_1} } \cdot  \omega \in \mathcal{M}, \quad \forall z \in\mathbb{C}.
\end{align}
Note that $\rho(\theta)=\oplus_\omega P(\omega)^{\frac{1}{\al}} \omega$ and
\begin{align}
\norm{\rho(\theta)}{L_1^\al(\mathcal{N}\subset\mathcal{M} )}&:=n^{\frac{1}{\al'}}\inf_{\si\in\mathcal{S(H)}} \norm{ (\si^{\oplus n})^{-\frac{1}{2\al'}} \rho(\theta) (\si^{\oplus n})^{-\frac{1}{2\al'}} }{\al} \\
&= n^{\frac{1}{\al'}}\inf_{\si\in\mathcal{S(H)}} \left( \sum_\omega P(\omega) \left\| \si^{\frac{1-\alpha}{2\alpha} }  \omega \si^{\frac{1-\alpha}{2\alpha}} \right\|_\alpha^\alpha \right)^{\frac{1}{\alpha}}, \label{eq:log_convex_norm2}
\end{align}
where $\frac{1}{\alpha} + \frac{1}{\alpha'} = 1$. Similarly, we have $\rho(0)=\oplus_\omega P(\omega)^{\frac{1}{\al_0}} \omega$,  $\rho(1)=\oplus_\omega P(\omega)^{\frac{1}{\al_1}} \omega$ and
\begin{align*}
&\norm{\rho(0)}{L_1^{\al_0}(\mathcal{N}\subset\mathcal{M} )}= n^{\frac{1}{\al_0'}}\inf_{\si\in\mathcal{S(H)}} \left( \sum_\omega P(\omega) \left\| \si^{\frac{1-\al_0}{2\al_0} }  \omega \si^{\frac{1-\al_0}{2\al_0}} \right\|_{\al_0}^{\al_0} \right)^{\frac{1}{\al_0}};\\
&\norm{\rho(1)}{L_1^{\al_1}(\mathcal{N}\subset\mathcal{M} )}= n^{\frac{1}{\al_1'}}\inf_{\si\in\mathcal{S(H)}} \left( \sum_\omega P(\omega) \left\| \si^{\frac{1-\al_1}{2\al_1} }  \omega \si^{\frac{1-\al_1}{2\al_1}} \right\|_{\al_1}^{\al_1} \right)^{\frac{1}{\al_1}}.
\end{align*}
Because of the complex interpolation relation $L_{1}^{\alpha}(\iN\subset\iM)=[L_{1}^{\alpha_0}(\iN\subset\iM),L_{1}^{\alpha_1}(\iN\subset\iM)]_\theta$ mentioned in \eqref{eq:inter_sandwiched} of Section~\ref{sec:interpolation},
we have the interpolation inequality \eqref{in}:
\begin{align} \label{eq:sand_Renyi1}
\norm{\rho(\theta)}{L_{1}^{\alpha}(\iN\subset\iM) }\le \left(\sup_{t\in\mathbb{R}}\norm{\rho( \mathrm{i}t )}{L_{1}^{\alpha_0}(\iN\subset\iM) }\right)^{1-\theta} \left(\sup_{t\in\mathbb{R}} \norm{\rho(1+\mathrm{i}t)}{L_{1}^{\alpha_1}(\iN\subset\iM) }\right)^\theta \pl.
\end{align}
Now we claim that for all $\al\geq 1$ and $t\in\mathbb{R}$,
\begin{align}
\norm{\rho(\mathrm{i} t)}{L_1^{\al}(\mathcal{N}\subset\mathcal{M} )} &\leq  \norm{\rho(0)}{L_1^{\al_0}(\mathcal{N}\subset\mathcal{M} )}; \label{eq:boundary1} \\
\norm{\rho(1+\mathrm{i} t)}{L_1^p(\mathcal{N}\subset\mathcal{M} )} &\leq  \norm{\rho(1)}{L_1^{\al_1}(\mathcal{N}\subset\mathcal{M} )}. \label{eq:boundary2}
\end{align}
Write $\displaystyle X(\omega,t)=P(\omega)^{ \frac{-\mathrm{i}t}{\alpha_0} + \frac{\mathrm{i}t}{\alpha_1}}$. Then, using H\"older inequality, it holds for all $p\geq 1$ and $t\in\mathbb{R}$ that,
\begin{align}
\norm{\rho(\mathrm{i} t)}{L_1^p(\mathcal{N}\subset\mathcal{M} )}
&= \inf_{\sigma_1, \sigma_2\in\mathcal{S(H)}} \left\| \bigoplus_{\omega} X(t,\omega) P(\omega)^{\frac{1}{\alpha_0} }\sigma_1^{ \frac{1-p}{2p}  } \omega \sigma_2^{ \frac{1-p}{2p}  } \right\|_{p}
\\&\leq \inf_{\sigma_1\in\mathcal{S(H)}} \left\| \bigoplus_{\omega}X(t,\omega) P(\omega)^{\frac{1}{2\alpha_0} } \sigma_1^{ \frac{1-p}{2p}  } \omega^\frac12 \right\|_{2p}
\inf_{\sigma_2\in\mathcal{S(H)}} \left\| \bigoplus_{\omega} P (\omega)^{\frac{1}{2\alpha_0}} \omega^\frac12 \sigma_2^{ \frac{1-p}{2p}  }  \right\|_{2p} \\
&= \inf_{\sigma_1\in\mathcal{S(H)}} \left\| \bigoplus_{\omega} P(\omega)^{\frac{1}{\alpha_0} } \sigma_1^{ \frac{1-p}{2p}  } \omega^\frac12 \sigma_1^{ \frac{1-p}{2p}  }\right\|_{p}^{\frac{1}{2}}
\inf_{\sigma_2\in\mathcal{S(H)}} \left\| \bigoplus_{\omega} P(\omega)^{\frac{1}{\alpha_0} } \sigma_2^{ \frac{1-p}{2p}  } \omega^\frac12 \sigma_2^{ \frac{1-p}{2p}  }\right\|_{p}^{\frac{1}{2}}\label{eq:X}\\
&= \norm{\rho(0)}{L_1^p(\mathcal{N}\subset\mathcal{M} )}^\frac12 \norm{\rho(0)}{L_1^p(\mathcal{N}\subset\mathcal{M} )}^\frac12= \norm{\rho(0)}{L_1^p(\mathcal{N}\subset\mathcal{M} )},
\end{align}
where equality \eqref{eq:X} holds since $X(t,\omega)\overline{X(t,\omega)}=1$. Then, the assertion in \eqref{eq:boundary2} follows similar argument.
Therefore, combining \eqref{eq:sand_Renyi1}, \eqref{eq:boundary1}, and \eqref{eq:boundary2}, we obtain that
\begin{align}
\norm{\rho(\theta)}{L_{1}^{\alpha}(\iN\subset\iM) }\le \norm{\rho(0)}{L_{1}^{\alpha_0}(\iN\subset\iM) }^{\theta}\norm{\rho(1)}{L_{1}^{\alpha_1}(\iN\subset\iM) }^{\theta}  \pl.
\end{align}
which yields the desired \eqref{eq:log_convex_norm1}. Hence, concavity $s\mapsto E_0^{\texttt{r},*}(s,P)$ for $s \in(-1,0)$ is proved.

 Next, we prove the concavity of $E_0^{*,\texttt{a}}(s,P)$ under the assumption  $\mathcal{H}$ is finite-dimensional. By the continuity in the probability distribution $P$ from Proposition~\ref{prop:I2}-\ref{I2-c}, it suffices to consider a finitely supported $P$ such that for each $\omega$, the probability $P(\omega)$ is rational. We can write $\displaystyle P(\omega)=\frac{n_\omega}{n}$ with some positive integers $n_\omega$ and $n=\sum_{\omega \in \mathscr{W} } n_\omega$.
	Given such a distribution $P$, we choose the following product state in $\mathcal{B(H)}^{\otimes n}$:
	\begin{align}  \rho:= &\underbrace{W_1\ten W_1\ten \cdots \ten W_1}_{n_1\text{-fold tensor}} \ten \underbrace{W_2\ten  \cdots \ten W_2}_{n_2\text{-fold tensor}} \ten  \cdots \ten \underbrace{W_k\ten  \cdots \ten W_k}_{n_k\text{-fold tensor}}\nonumber
	\\= &\big(W_1^{\ten n_1}\big)\ten \big(W_2^{\ten n_2}\big)\ten \cdots \ten \big(W_k^{\ten n_k}\big), \label{om}
	\end{align}
	where $\{W_1, W_2, \ldots, W_k \} \subseteq \mathscr{W}$. We have \begin{align}I_{\al}^{\texttt{a},*}(P,\mathscr{W})&=\inf_{\si\in\mathcal{S(H)}}\sum_{ \omega \in \mathscr{W} } P(\omega) D^*_\al( \omega  ||\si) \nonumber
	\\ &=\inf_{\si\in\mathcal{S(H)}}\sum_{ \omega \in \mathscr{W} } P(\omega)\al'\log \norm{\si^{-\frac{1}{2\al'}} \omega \si^{-\frac{1}{2\al'}}}{\al}\nonumber
	\\ &=\frac{\al'}{n}\inf_{\si\in\mathcal{S(H)}}\log \norm{\si^{-\frac{1}{2\al'}}W_1 \si^{-\frac{1}{2\al'}}}{\al}^{n_1}\cdot \norm{\si^{-\frac{1}{2\al'}}W_2 \si^{-\frac{1}{2\al'}}}{\al}^{n_2}\cdots \norm{\si^{-\frac{1}{2\al'}}W_k \si^{-\frac{1}{2\al'}}}{\al}^{n_k}\nonumber
	\\ &=\frac{\al'}{n}\log \inf_{\si\in\mathcal{S(H)}} \norm{(\si^{\otimes n})^{-\frac{1}{2\al'}}\rho (\si^{\otimes n} )^{-\frac{1}{2\al'}}}{\al} \label{c}
	\end{align}
	where $\rho$ is the density chosen in \eqref{om}, and $\alpha'$ is the conjugate of $\alpha$, i.e.~$\frac1{\alpha} + \frac1{\alpha'} = 1$.
	Similar as for $E_0^{\texttt{r},*}(s,P)$ it suffices to prove the following interpolation-type inequality: for all $1\le \alpha_0,\al_1\le \infty$, $\frac{1}{\alpha} = \frac{1-\theta}{\alpha_0} + \frac{\theta}{\alpha_1}$, $\theta \in [0,1]$,
	\begin{align}
	\begin{split}
	&\quad\inf_{ \sigma\in\mathcal{S(H)}} \left\| (\sigma^{\otimes n} )^{-\frac{1}{2\alpha'}} \rho (\sigma^{\otimes n} )^{-\frac{1}{2\alpha'}} \right\|_{\alpha} \\
	&\leq
	\inf_{ \sigma\in\mathcal{S(H)}} \left\| (\sigma^{\otimes n} )^{-\frac{1}{2\alpha'_0}} 	\rho (\sigma^{\otimes n} )^{-\frac{1}{2\alpha'_0}} \right\|_{\alpha_0}^{1-\theta}
	\inf_{ \sigma\in\mathcal{S(H)}} \left\| (\sigma^{\otimes n} )^{-\frac{1}{2\alpha'_1}} 	\rho (\sigma^{\otimes n} )^{-\frac{1}{2\alpha'_1}} \right\|_{\alpha_1}^\theta.
	\end{split}
	\end{align}
	Note that the product states $\si^{\ten n}$ does not form a convex set; hence it is not the state space of a subalgebra. So for this case, the argument using amalgamated $L_p$ spaces does not apply. Instead, we provide a proof using direct interpolation. First,
	\[ \inf_{ \sigma\in\mathcal{S(H)}} \left\| (\sigma^{\otimes n} )^{-\frac{1}{2\alpha'}} \rho (\sigma^{\otimes n} )^{-\frac{1}{2\alpha'}} \right\|_{\alpha}=\inf_{ \norm{\tau}{2\alpha'}=1} \left\| (\tau^{\otimes n} )^{-1} \rho (\tau^{\otimes n} )^{-1} \right\|_{\alpha}\ .\]
	where the infimum is taken over all positive $\tau$ with $\norm{\tau}{2\al'}=1$. Given $1\le \alpha_0,\al_1\le \infty$ and $\epsilon >0$, we choose $\tau_0$ and $\tau_1$ such that $\norm{\tau_0}{2\al_0'}=\norm{\tau_1}{2\al_1'}=1$ and
	\begin{align}
	&\left\| (\tau_0^{\otimes n} )^{-1} \rho (\tau_0^{\otimes n} )^{-1} \right\|_{\alpha_0}\le (1+\epsilon)\inf_{ \|{\tau}\|_{2\al_0'}=1} \left\| (\tau^{\otimes n} )^{-1} \rho (\tau^{\otimes n} )^{-1} \right\|_{\alpha_0},\\
	&\left\| (\tau_1^{\otimes n} )^{-1} \rho (\tau_1^{\otimes n} )^{-1} \right\|_{\alpha_1}\le (1+\epsilon)\inf_{ \|{\tau}\|_{2\al_1'}=1} \left\| (\tau^{\otimes n} )^{-1} \rho (\tau^{\otimes n} )^{-1} \right\|_{\alpha_1}.
	\end{align}
Because $\mathcal{H}$ is finite-dimensional, we can assume there is a $\delta>0$ such that $\tau_0,\tau_1\ge \delta \mathds{1}$. Using Devinatz’s factorization theorem \cite{devinatz61} (see also Pisier’s paper \cite[Theorem 3.2]{pisier92}), there exists an operator valued analytic function $w:\{z\,|0\le \textsf{Re} (z)\le 1\} \to \mathcal{B(H)}$ such that $w(z)$ is invertible for all $z$ with $z\mapsto w(z)^{-1}$ is bounded and analytic and moreover,
	\[ w(\mathrm{i} t)w(\mathrm{i} t)^*=\tau_0^2,  \quad w(1+\mathrm{i} t)w(1+\mathrm{i} t)^*=\tau_1^2, \quad \forall t\in\mathbb{R}.\]
	Then \begin{align}
	&\|w(\mathrm{i} t)\|_{2\alpha_0'}=\|w(\mathrm{i} t)w(\mathrm{i} t)^*\|_{\alpha_0'}^{\frac{1}{2}}=\|\tau_0^2\|_{\alpha_0'}^{\frac{1}{2}}=\|\tau_0\|_{2\alpha_0'}=1,\\
	&\|w(1+\mathrm{i} t)\|_{2\alpha_1'}=\|w(1+\mathrm{i} t)w(1+\mathrm{i} t)^*\|_{\alpha_1'}^{\frac{1}{2}}=\|\tau_1^2\|_{\alpha_1'}^{\frac{1}{2}}=\|\tau_1\|_{2\alpha_1'}=1,
	\end{align}
	which implies
	\begin{align}
	\|w(\theta)\|_{2\alpha' }\leq \left( \sup_{t} \|w(\mathrm{i} t)\|_{2\alpha_0'} \right)^{1-\theta} \left( \sup_t \|w(1+\mathrm{i} t)\|_{2\alpha_1'} \right)^{\theta} = 1
	\end{align}
	by interpolation inequality. Next, consider the analytic function
	\[f(z)=\rho^{\frac{1}{2}}\Big(w(z)\ten w(z)\ten \cdots \ten w(z)\Big)^{-1}, \quad z \in \{z\,|0\le \textsf{Re} (z)\le 1\}.
	\]
	Note that for all $t \in \mathbb{R}$,
	\begin{align}
	&\norm{f(\mathrm{i} t)}{
		2\al_0}=\left\|( w^*(\mathrm{i} t)^{\otimes n} )^{-1} \rho (w(\mathrm{i} t)^{\otimes n} )^{-1} \right\|_{\alpha_0}=\left\| (\tau_0^{\otimes n} )^{-1} \rho (\tau_0^{\otimes n} )^{-1} \right\|_{\alpha_0} \ , \\ &\ \norm{f(1+\mathrm{i} t)}{
		2\al_1}=\left\| (w^*(1+\mathrm{i} t)^{\otimes n} )^{-1} \rho (w(1+\mathrm{i} t)^{\otimes n} )^{-1}\right\|_{\alpha_1}=\left\| (\tau_1^{\otimes n} )^{-1} \rho (\tau_1^{\otimes n} )^{-1} \right\|_{\alpha_1}\pl,
	\end{align}
	because the polar decomposition $w(\mathrm{i} t)=u(t)\tau_0^{\frac{1}{2}}, w(1+\mathrm{i} t)=v(t)\tau_0^{\frac{1}{2}}$ for some unitary function $u(t),v(t)$.
	Then by interpolation,
	\begin{align}
	\norm{f(\theta)}{2\alpha }^2&\le \Big(\sup_t\norm{f(\mathrm{i} t)}{2\alpha_0}^2\Big)^{1-\theta} \Big(\sup_t\norm{f(1+\mathrm{i} t)}{
		2\alpha_1}^2\Big)^{\theta}
	\\
	&= \left\| (\tau_0^{\otimes n} )^{-1} \rho (\tau_0^{\otimes n} )^{-1} \right\|_{\alpha_0}^{1-\theta}\left\| (\tau_1^{\otimes n} )^{-1} \rho (\tau_1^{\otimes n} )^{-1} \right\|_{\alpha_1}^{\theta}
	\\
	&\le (1+\epsilon)^2\Big(\inf_{ \norm{\tau}{2\alpha_0'}=1} \left\| (\tau^{\otimes n} )^{-1} \rho (\tau^{\otimes n} )^{-1} \right\|_{\alpha_0}\Big)^{1-\theta}\Big(\inf_{ \norm{\tau}{2\alpha_1'}=1} \left\| (\tau^{\otimes n} )^{-1} \rho (\tau^{\otimes n} )^{-1} \right\|_{\alpha_1}\Big)^{\theta}.
	\end{align}
Note that $\norm{w(\theta)^*}{2\al' }\le 1$. Choosing $\tau=\frac{|w(\theta)|}{\norm{w(\theta)}{2\al' }}$,
	\begin{align}
	\norm{f(\theta)}{2\alpha }^2 &=
	\norm{f(\theta) f(\theta)^*}{\al } \\
	&=\left\| (w^*(\theta)^{\otimes n} )^{-1} \rho (w(\theta)^{\otimes n} )^{-1} \right\|_{\alpha }\\
	&=\left\| (|w(\theta)|^{\otimes n} )^{-1} \rho (|w(\theta)|^{\otimes n} )^{-1} \right\|_{\alpha }\\
	&=\left\| (\tau^{\otimes n} )^{-1} \rho (\tau^{\otimes n} )^{-1} \right\|_{\alpha } \norm{w(\theta)}{2\al' }^{-2} \\
	&\geq\left\| (\tau^{\otimes n} )^{-1} \rho (\tau^{\otimes n} )^{-1} \right\|_{\alpha }  \\
	&\ge \inf_{\norm{\tau}{2\al'}=1} \left\| (\tau^{\otimes n} )^{-1} \rho (\tau^{\otimes n} )^{-1} \right\|_{\alpha }.
	\end{align}
	Taking $\epsilon \to 0$, we obtain the desired interpolation inequality.
\end{proof}


Next, we move on to prove the concavity of concavity of  auxiliary functions of Petz form, i.e.~ $E_0^\texttt{r}(s,P)$ and $E_0^\texttt{a}(s,P)$.
The main ingredients are the noncommutative Sibson's identity proved in Proposition~\ref{prop:Sibson} in Section~\ref{sec:interpolation} and Devinatz’s factorization theorem \cite{devinatz61}.

\begin{theo}[Concavity of Petz Form in Order] \label{theo:concave_Petz}
Let $\mathscr{W} \subset \mathcal{S(H)}$ and $P\in\mathscr{P(W)}$ be a probability mass function.
\begin{enumerate}[(a)]
\item 
The map $s\mapsto E_0^{\textnormal{\texttt{r}}}(s,P)$ is concave on $(-1,0)$.
\item\label{concave_Petz-b} If $\mathcal{H}$ is finite-dimensional, then $s\mapsto E_0^{\textnormal{\texttt{a}}}(s,P)$ is concave on $(-1,0)$.
\end{enumerate}
\end{theo}

\begin{proof}[Proof of Theorem~\ref{theo:concave_Petz}]
	Note that
	\begin{align}
	I_\alpha^{\texttt{r}}(P,\mathscr{W}) &= \inf_{\si \in \mathcal{S(H)}} \frac{1}{\alpha-1} \log \left(\sum_\omega P(\omega) \Tr\left[ \omega^{\alpha} \sigma^{1-\alpha} \right] \right) \\
	&= \inf_{\si \in \mathcal{S(H)}} \frac{1}{\alpha-1} \log \left\| \bigoplus_{\omega} P^{\frac12} (\omega)  \omega^{\frac{\alpha}{2}}  \sigma^{\frac{1-\alpha}{2}} \right\|_2.
	\end{align}
	Hence, the concavity of $s\mapsto E_0(s,P)$ for $s \in(-1,0)$ is equivalent to showing that for all $\alpha_0,\alpha_1 \geq 1$ with $\frac{1}{\alpha} = \frac{1-\theta}{\alpha_0} + \frac{\theta}{\alpha_1}$, $\theta \in [0,1]$, and $\sigma_0,\sigma_1 \in \mathcal{S(H)}$,
	\begin{align}
	\inf_{\sigma \in \mathcal{S(H)}}  \Tr\left[ \Big(\sum_\omega P(\omega)\omega^{\alpha}\Big) \sigma^{1-\alpha} \right]^{\frac{1}{\al}}
	\leq
	\left\| \bigoplus_{\omega} P^{\frac12} (\omega)  \omega^{\frac{\alpha_0}{2}}  \sigma_0^{\frac{1-\alpha_0}{2}} \right\|_2^{\frac{1-\theta}{\alpha_0}}
	\left\| \bigoplus_{\omega} P^{\frac12} (\omega)  \omega^{\frac{\alpha_1}{2}}  \sigma_1^{\frac{1-\alpha_1}{2}} \right\|_2^{\frac{\theta}{\alpha_1}}.
	\end{align}
	To prove this, we employ a similar argument as our proof of Proposition~\ref{theo:inter_Petz} in Section~\ref{sec:interpolation}.
	Denote $\gamma = \frac{\alpha(1-\theta)}{\alpha_0}$ and $1-\gamma = \frac{\alpha\theta}{\alpha_1}$. Let us first consider the case $\gamma = \frac12$ and $\alpha = \frac12(\alpha_0 +
	\alpha_1)$. Then, we start with Cauchy-Schwartz inequality:
	\begin{align}
	\left\| \bigoplus_{\omega} P^{\frac12} (\omega)  \omega^{\frac{\alpha_0}{2}}  \sigma_0^{\frac{1-\alpha_0}{2}} \right\|_2
	\left\| \bigoplus_{\omega} P^{\frac12} (\omega)  \omega^{\frac{\alpha_1}{2}}  \sigma_1^{\frac{1-\alpha_1}{2}} \right\|_2
	&\geq\left\| \bigoplus_\omega P(\omega) \sigma_0^{\frac{1-\alpha_0}{2}}\omega^{\frac{\alpha_0+\alpha_1}{2} } \sigma_1^{\frac{1-\alpha_1}{2}} \right\|_1
	\\
	&\geq\left\| \mathbb{E}\left[ \bigoplus_\omega P(\omega) \sigma_0^{\frac{1-\alpha_0}{2}}\omega^{\frac{\alpha_0+\alpha_1}{2} } \sigma_1^{\frac{1-\alpha_1}{2}}  \right] \right\|_1 \label{eq:Petz_Renyi3}
	\\
	&= \left\| \Big(\bigoplus_\omega \sigma_0^{\frac{1-\alpha_0}{2}}\Big)  \mathbb{E}\left[\bigoplus_\omega P(\omega) \omega^\al \right]\Big(\bigoplus_\omega \sigma_1^{\frac{1-\alpha_1}{2}} \Big)\right\|_1 \label{eq:Petz_Renyi4}
\\ &= \left\| \sigma_0^{\frac{1-\alpha_0}{2}} \Big(\sum_\omega P(\omega) \omega^\al\Big)\sigma_1^{\frac{1-\alpha_1}{2}} \right\|_1\\
&\geq \left\|\sum_{\omega} P(\omega) \omega^{ \alpha } \right\|_{\frac{1}{\alpha}} \label{eq:Petz_Renyi1} \\
	&= \inf_{\si \in \mathcal{S(H)}} \Tr\left[ \Big(\sum_\omega P(\omega) \omega^{\alpha}\Big) \sigma^{1-\alpha} \right]. \label{eq:Petz_Renyi2}
	\end{align}
Here, we denote $\mathbb{E}$ as the conditional expectation from the algebra $\mathcal{M}:= \oplus_{x\in \mathcal{X}} \mathcal{B(H)}$ to the subalgebra $\mathcal{N}:= \mathds{1}_\mathcal{X}\ten\mathcal{B(H)}$ for $\mathcal{X} := \texttt{supp}(P)$, i.e.
\[\mathbb{E}\left[\bigoplus_{x \in \mathcal{X}} \rho_x \right] =1_{\mathcal{X}}\ten\left(\frac{1}{|\mathcal{X}|}\sum_{x\in\mathcal{X}}\rho_x\right)\pl, \quad \forall \rho_x \in \mathcal{B(H)}.
\]
Hence, inequality~\eqref{eq:Petz_Renyi3} follows from the contraction of $1$-norm under $\mathbb{E}$.
In Eq.~\eqref{eq:Petz_Renyi4}, we employ the module property of $\mathbb{E}$, i.e.~
\begin{align}
\mathbb{E}\left(axb \right) = a \mathbb{E}(x) b, \quad \forall a,b\in\mathcal{N},\; x \in \mathcal{M}.
\end{align}
The last inequality \eqref{eq:Petz_Renyi1} is owing to the H{\"{o}}lder inequality for $\al = 1 + \frac{\al_0-1}{2} + \frac{\al_1-1}{2}$ and the fact that $\sigma_0$ and $\sigma_1$ are density operators, i.e.
	\begin{align}
	\left\|  \sum_\omega P(\omega) \omega^{\alpha} \right\|_{\frac{1}{\al}}
	&\leq \left\| \sigma_1^{\frac{\al_1-1}{2}} \right\|_{\frac{2}{\al_1-1}} \left\| \sigma_1^{\frac{1-\al_1}{2} } \Big(\sum_\omega P(\omega) \omega^{\alpha}\Big)\sigma_0^{\frac{1-\al_0}{2}} \right\|_1 \left\| \sigma_0^{\frac{\al_0-1}{2}} \right\|_{\frac{2}{\al_0-1}} \\
	&\leq \left\| \sigma_1^{\frac{1-\al_1}{2} } \Big(\sum_\omega P(\omega) \omega^{\alpha}\Big)\sigma_0^{\frac{1-\al_0}{2}} \right\|_1\pl.
	\end{align}
	In the last equality \eqref{eq:Petz_Renyi2}, we apply the noncommutative Sibson's identity given in Proposition~\ref{prop:Sibson}.
	
This proves the inequality for $\gamma = \frac12$. Using induction, we obtain the inequality for $2^n$-partition points $\al = k 2^{-n}(\alpha_1-\alpha_0) + \alpha_0$ for all $k,n\in\mathbb{N}$, $0\leq k \leq 2^n$. The case for general $\alpha$ follows from the continuity.

\medskip

Next, we prove the concavity of $s\mapsto E_0^{\texttt{a}}(s,P)$.
In the following, we assume that the Hilbert space $\mathcal{H}$ is finite-dimensional.
Using the construction in Eqs.~\eqref{om}, it suffices to prove the following interpolation type inequality:  for all $1\le \alpha_0,\al_1\le \infty$, $\frac{1}{\alpha} = \frac{1-\theta}{\alpha_0} + \frac{\theta}{\alpha_1} $, $\theta \in [0,1]$,
\begin{align}
\begin{split} \label{eq:inter_3}
\quad\inf_{ \sigma\in\mathcal{S(H)}} \Tr\left[ \rho^\alpha (\sigma^{\otimes n} )^{1-\alpha} \right]^\frac{1}{\alpha}
\leq
\inf_{ \sigma\in\mathcal{S(H)}} \Tr\left[ \rho^{\alpha_0} (\sigma^{\otimes n} )^{1-\alpha_0} \right]^{\frac{1-\theta}{\alpha_0}} \cdot
\inf_{ \sigma\in\mathcal{S(H)}} \Tr\left[ \rho^{\alpha_1} (\sigma^{\otimes n} )^{1-\alpha_1} \right]^{\frac{\theta}{\alpha_1}}.
\end{split}
\end{align}
Note that
\begin{align}
\inf_{ \sigma\in\mathcal{S(H)}} \Tr\left[ \rho^\alpha (\sigma^{\otimes n} )^{1-\alpha} \right]
=\inf_{ \sigma\in\mathcal{S(H)}} \left\| (\sigma^{\otimes n})^{\frac{1-\alpha}{2}} \rho^\frac{\alpha}{2}  \right\|_2^2 = \inf_{ \|\tau\|_{\frac{2}{\alpha-1}} = 1 } \left\| ( \tau^{\otimes n})^{-1} \rho^\frac{\alpha}{2}  \right\|_2^2,
\end{align}
where the infimum in the last line is taken over all invertible positive $\tau$ with $\|\tau\|_{\frac{2}{\alpha-1}} = 1$.
Letting $\lambda = \frac{\theta \alpha}{\alpha_1} \in [0,1]$, inequality \eqref{eq:inter_3} can be rewritten as
\begin{align}
\begin{split}
\quad\inf_{ \|\tau\|_{\frac{2}{\alpha-1}} = 1 } \left\| ( \tau^{\otimes n})^{-1} \rho^\frac{\alpha}{2}  \right\|_2 \leq
\inf_{ \|\tau\|_{\frac{2}{\alpha_0-1}} = 1 } \left\| ( \tau^{\otimes n})^{-1} \rho^\frac{\alpha_0}{2}  \right\|_2^{1-\lambda} \cdot
\inf_{ \|\tau\|_{\frac{2}{\alpha_1-1}} = 1 } \left\| ( \tau^{\otimes n})^{-1} \rho^\frac{\alpha_1}{2}  \right\|_2^{\lambda}.
\end{split}
\end{align}
Given $1\le \alpha_0,\al_1\le \infty$ and $\epsilon >0$, we choose $\tau_0$ and $\tau_1$ such that $\|\tau_0 \|_{\frac{2}{\alpha_0-1}} = \|\tau_1 \|_{\frac{2}{\alpha_0-1}} =1$ and
\begin{align}
&\left\| ( \tau_0^{\otimes n})^{-1} \rho^\frac{\alpha_0}{2}  \right\|_2 \le (1+\epsilon)\inf_{ \|\tau\|_{\frac{2}{\alpha_0-1}} = 1  } \left\| ( \tau^{\otimes n})^{-1} \rho^\frac{\alpha_0}{2}  \right\|_2, \\
&\left\| ( \tau_1^{\otimes n})^{-1} \rho^\frac{\alpha_1}{2}  \right\|_2 \le (1+\epsilon)\inf_{ \|\tau\|_{\frac{2}{\alpha_1-1}} = 1  } \left\| ( \tau^{\otimes n})^{-1} \rho^\frac{\alpha_1}{2}  \right\|_2.
\end{align}
Using Devinatz’s factorization theorem again as in the proof of Theorem~\ref{theo:concave_*}, there exists an operator valued analytic function $w:\{z\,|0\le \textsf{Re} (z)\le 1\} \to \mathcal{B(H)}$ such that $w(z)$ is invertible for all $z$ and
\[ w(\mathrm{i} t)w(\mathrm{i} t)^*=\tau_0^2\pl,  w(1+\mathrm{i} t)w(1+\mathrm{i} t)^*=\tau_1^2\pl, \quad \forall t\in\mathbb{R} .\]
Then \begin{align}
&\norm{w(\mathrm{i} t)}{ \frac{2}{\alpha_0-1} } = \norm{w(\mathrm{i} t)w(\mathrm{i} t)^*}{ \frac{1}{\alpha_0-1} }^{\frac{1}{2}} = \norm{\tau_0^2}{ \frac{1}{\alpha_0-1} }^{\frac{1}{2}} = \norm{\tau_0}{ \frac{2}{\alpha_0-1} } = 1\\
&\norm{w(1+\mathrm{i} t)}{ \frac{2}{\alpha_1-1} } = \norm{w(1+\mathrm{i} t)w(1+\mathrm{i} t)^*}{ \frac{1}{\alpha_1-1} }^{\frac{1}{2}} = \norm{\tau_1^2}{ \frac{1}{\alpha_1-1} }^{\frac{1}{2}} = \norm{\tau_1}{ \frac{2}{\alpha_1-1} } = 1.
\end{align}
Using $\frac{1}{\alpha} = \frac{1-\theta}{\alpha_0} + \frac{\theta}{\alpha_1}$ and $\lambda = \frac{\theta \alpha}{\alpha_1}$, one can verify that
\begin{align}
 \frac{\alpha_0-1}{2}(1-\lambda) + \frac{\alpha_1-1}{2}\lambda
= \frac{\alpha-1}{2}.
\end{align}
Note that for $1<\al<\infty$, $0<\frac{2}{\alpha-1}\le \infty$. The interpolation inequality also holds for $L_p, 0<p<\infty$ \cite[Lemma 2.5]{richard}.
We have
\begin{align}
\norm{w(\lambda)}{\frac{2}{\alpha-1}}\le
\left( \sup_t \norm{w(\mathrm{i} t)}{ \frac{2}{\alpha_0-1} } \right)^{1-\lambda}
\left( \sup_t \norm{w(1+\mathrm{i} t)}{ \frac{2}{\alpha_1-1} } \right)^{\lambda} =
1.
\end{align}
Next, consider the analytic function
\[f(z)=\rho^{\frac{\alpha_0}{2} + \frac{\alpha_1-\alpha_0}{2}z} \Big(w(z)\ten w(z)\ten \cdots \ten w(z)\Big)^{-1}, \quad z \in \{z\,|0\le \textsf{Re} (z)\le 1\}.\]
Note that for all $t \in \mathbb{R}$,
\begin{align}
&\norm{f(\mathrm{i} t)}{
	2}=\left\| \rho^{\frac{\alpha_0}{2} + \frac{\alpha_1-\alpha_0}{2}(\mathrm{i} t)} w(\mathrm{i} t)^{\otimes n} )^{-1} \right\|_{2} =\left\| (\tau_0^{\otimes n} )^{-1} \rho^{\frac{\alpha_0}{2} }  \right\|_{2}, \\
&\norm{f(1+\mathrm{i} t)}{
	2}=\left\| \rho^{\frac{\alpha_0}{2} + \frac{\alpha_1-\alpha_0}{2}(1+\mathrm{i} t)} \right\|_{2} =\left\| (\tau_1^{\otimes n} )^{-1} \rho^{\frac{\alpha_1}{2} }(w(1+\mathrm{i} t)^{\otimes n} )^{-1}   \right\|_{2},
\end{align}
because the polar decomposition $w(\mathrm{i} t)=u(t)\si_0^{\frac{1}{2}}, w(1+\mathrm{i} t)=v(t)\si_0^{\frac{1}{2}}$ for some unitary function $u(t),v(t)$.
By interpolation inequality again,
\begin{align}
\norm{f(\lambda)}{2}&\le \left(\sup_{t\in\mathbb{R}}\norm{f(\mathrm{i} t)}{2}\right)^{1-\lambda} \Big(\sup_{t\in\mathbb{R}}\norm{f(1+\mathrm{i} t)}{
	2}\Big)^{\lambda}
\\
&= \left\| ( \tau_0^{\otimes n})^{-1} \rho^\frac{\alpha_0}{2}  \right\|_2^{1-\lambda} \left\| ( \tau_1^{\otimes n})^{-1} \rho^\frac{\alpha_1}{2}  \right\|_2^{\lambda}
\\
&\le (1+\epsilon)^2 \left(\inf_{ \|\tau\|_{\frac{2}{\alpha_0-1}} = 1  } \left\| ( \tau^{\otimes n})^{-1} \rho^\frac{\alpha_0}{2}  \right\|_2 \right)^{1-\lambda}\left(\inf_{ \|\tau\|_{\frac{2}{\alpha_1-1}} = 1  } \left\| ( \tau^{\otimes n})^{-1} \rho^\frac{\alpha_1}{2}  \right\|_2 \right)^{\lambda}.
\end{align}
On the other hand, by $\norm{w(\lambda)}{ \frac{2}{\alpha-1}  }\le 1$ and choosing $\tau=\frac{|w(\lambda)|}{\norm{w(\lambda)}{ \frac{2}{\alpha-1}  }}$,
\begin{align}
\norm{f(\lambda)}{2}&=\left\|  \rho^{ \frac{\alpha_0}{2} + \frac{\alpha_1-\alpha_0}{2}\lambda }(w(\lambda)^{\otimes n} )^{-1} \right\|_{2}\\
&=\left\| \rho^{ \frac{\alpha}{2} } (w(\lambda)^{\otimes n} )^{-1} \right\|_{2}\\
&=\left\|  \rho^{ \frac{\alpha}{2} } (|w(\lambda)|^{\otimes n} )^{-1}\right\|_{2}\\
&=\left\|  \rho^\frac{\alpha}{2} (\tau^{\otimes n} )^{-1}\right\|_{2} \norm{w(\lambda)}{ \frac{2}{\alpha-1} }^{-1} \\
&\geq\left\|  \rho^\frac{\alpha}{2}(\tau^{\otimes n} )^{-1} \right\|_{2} \\
&\ge \inf_{\norm{\tau}{\frac{2}{\alpha-1}}=1} \left\| \rho^\frac{\alpha}{2}(\tau^{\otimes n} )^{-1}  \right\|_{2}.
\end{align}
Taking $\epsilon \to 0$, we obtain the desired interpolation inequality.
\end{proof}
\begin{remark} \label{remark}
	The above concavity remains true if $\mathcal{B(H)}$ is replaced by tracial von Neumann algebras. In particular, the argument $E_0^{\texttt{r},*}$ and $E_0^{\texttt{r}}$ extends for all semifinite von Neumann algebra, and the argument $E_0^{\texttt{a},*}$ and $E_0^{\texttt{a}}$ works for all finite von Neumann algebra. Here the obstruction to extends the concavity of $E_0^{\texttt{a}}$ to the infinite-dimensional (semifinie) case is that Devinatz’s factorization theorem \cite{devinatz61} in our setting requires densities $\tau_0,\tau_1\ge \delta \mathds{1}$ for some $\delta > 0$. However, in the infinite dimensions, a density is bounded from below if and only if it is finite rank (its supported has finite trace). It is not known to us that the auxiliary functions can be approximated by finite rank densities.
\end{remark}


Finally, we apply the established concavity properties in Theorem~\ref{theo:concave_*} and \ref{theo:concave_Petz} and the equicontinuity of R\'enyi and Augustin information in prior input distributions (Propositions~\ref{prop:I2_R} and \ref{prop:I2}) to prove the following joint continuity.

\begin{theo}
	[Joint Continuity for Auxiliary Functions] \label{theo:joint_cont}
	Let $\mathscr{W} \subset \mathcal{S(H)}$.
	Assume $C_{\frac{1}{1+z},\mathscr{W}} < \infty$ for some $z > -1$.
	The following holds.
	\begin{enumerate}[(a)]
		\item (Petz) The map $(s,P)\mapsto E_0^{\textnormal{\texttt{r}}}(s,P)$ is jointly continuous on $[\max\{0,z\},\infty] \times \mathscr{P(W)}$, and $(s,P)\mapsto E_0^{\textnormal{\texttt{a}}}(s,P)$ is jointly continuous on $[z, 0) \times \mathscr{P(W)}$ and $(0,\infty] \times \mathscr{P(W)}$.
		
		\item (Sandwiched) The maps $(s,P)\mapsto E_0^{\textnormal{\texttt{r}}, *}(s,P)$ and $(s,P)\mapsto E_0^{\textnormal{\texttt{a}},*}(s,P)$ are jointly continuous on $[z,0)\times \mathscr{P(W)}$ and $(0,1] \times \mathscr{P(W)}$, respectively.		

		\item (Log-Euclidean) The maps $(s,P)\mapsto E_0^{\textnormal{\texttt{r}}, \flat}(s,P)$ and $(s,P)\mapsto E_0^{\textnormal{\texttt{a}},\flat}(s,P)$ are jointly continuous on $[z,0)\times \mathscr{P(W)}$ and $(0,\infty] \times \mathscr{P(W)}$, respectively.		
	\end{enumerate}
	The assertions for $E_0^\textnormal{\texttt{a}}$ and $E_0^{\textnormal{\texttt{a}},*}$ require an additional assumption of  $|\mathcal{H}| < \infty$.
\end{theo}

\begin{remark}
	Note that $z = -1$ if $|\mathcal{H}| < \infty$.
	The technical assumption of $|\mathcal{H}| < \infty$ is because of Theorems~\ref{theo:concave_*}-\ref{concave_*b} and \ref{theo:concave_Petz}-\ref{concave_Petz-b}.
	We conjecture that such the condition could be removed.
\end{remark}

\begin{remark}
	The concavity at $s=0$ is excluded in Theorem~\ref{theo:joint_cont} (except for $E_0^\texttt{r}$) since we do not know if the associated auxiliary functions are differentiable at $s=0$.
	We leave this as an open problem in Section~\ref{sec:conclusions}.
	Nevertheless, the regions of interest are either $s\in(-1,0)$ or $s \in (0,\infty)$ as they correspond to the \emph{strong converse regime} and the \emph{error exponent region} of classical-quantum channel coding and classical data compression with quantum side information as we will discuss in Section~\ref{sec:applications} below.
\end{remark}

\begin{proof}[Proof of Theorem~\ref{theo:joint_cont}]
	The concavity of $s\mapsto E_{0}^{\texttt{r},(t)}(s,P)$ and $s\mapsto E_{0}^{\texttt{a},(t)}(s,P)$ in $[0,\infty]$ obtained in   Corollary~B.2, and Proposition~B.5 of \cite{MO17} already imply that they are also continuous in $s \in [0,\infty]$.
	On the other hand, the established Theorems~\ref{theo:concave_*} and \ref{theo:concave_Petz} imply that they are continuous in $s \in (-1,0)$.
	Recalling the definitions given in Eq.~\eqref{eq:E0_c1} and \eqref{eq:E0_c2}, one of the assertions follows from the continuity in $s$ and the equicontinuity in $P$ proven in	Proposition~\ref{prop:I2_R}-\ref{I2_R-c} and Proposition~\ref{prop:I2}-\ref{I2-c}.
\end{proof}

\section{Applications of Auxiliary Functions in Error Exponent Analysis} \label{sec:applications}

Before commencing the section, we first introduce the two relevant information-processing tasks---the classical-quantum (c-q) channel coding and the classical source coding with quantum side information (QSI).
Then, we explain the roles of the auxiliary functions in error exponent analysis in these tasks.

\subsection{Quantum Information-Processing Tasks}
In the problems of c-q channel coding, the aim is to transmit classical information through a c-q channel with some coding strategy.
The classical information is represented by the messages in a finite message set, which we denote by $\mathcal{I}$. An ($n$-block) \emph{encoder} is a map from the message set to an $n$-fold input alphabet $\mathcal{X}$, i.e.~$f_n:\mathcal{I}\to \mathcal{X}^n$, such that each message $m\in\mathcal{I}$ is encoded to a codeword $\mathbf{x}^n(m) :=   x_1(m) x_2(m) \ldots x_n(m) \in\mathcal{X}^n$.
A classical-quantum (c-q) channel $\mathscr{W}:x\mapsto W_x$ is a map from symbols in the input alphabet $\mathcal{X}$ to a density operator on the output alphabet, which is conventionally modeled by some Hilbert space $B\equiv \mathcal{H}_B$.
Moreover, $n$-blocklength codeword $\mathbf{x}^n(m)$ through the c-q channel will be mapped to a product state:
\begin{align} \label{eq:W^n}
\mathscr{W}: \mathbf{x}^n(m) \mapsto
W_{\mathbf{x}^n(m)}^{\otimes n} = W_{x_1(m)} \otimes W_{x_2(m)} \otimes \cdots \otimes W_{x_n(m)} \in \mathcal{S}(B)^{\otimes n}.
\end{align}

The \emph{decoder} $\mathcal{D}_n$ is described by a positive operator-valued measurement (POVM) $\Pi = \{\Pi_{1},\ldots, \Pi_{|\mathcal{I}|} \}$ on $\mathcal{H}^{\otimes n}$, where $\Pi_{i} \geq 0$ and $\sum_{i=1}^{|\mathcal{I}|} \Pi_{i} = \mathds{1}$. The pair $(\mathcal{E}_n, \mathcal{D}_n) =: \mathcal{C}_n$ is called an $(n,R)$-\emph{code} with \emph{transmission rate} $R = \frac1n \log |\mathcal{C}_n| = \frac1n \log |\mathcal{I}|$.  The error probability of  sending a message $m$ with the code $ \mathcal{C}_n$ is $\eps_m(\mathcal{C}_n) :=  1- \Tr\left(\Pi_{m} W_{\mathbf{x}^n(m)}^{\otimes n}\right)$. The \emph{average} error probability is defined by $\bar{\eps}_\text{c}(\mathcal{C}_n) = \frac{1}{|\mathcal{I}|} \sum_{m\in\mathcal{I}} \eps_m(\mathcal{C}_n)$.
We denote by $\eps^\star_\text{c}\left(n, R \right)$ the {minimum} average probability of error among all the channel coding strategies with a blocklength $n$ and transmission rate $R$, i.e.
\begin{align}
\eps_\text{c}^\star (n,R) := \inf \{ \bar{\eps}_\text{c}(\mathcal{C}_n) : \mathcal{C}_n \text{ is an }(n,R)\text{-code}\},
\end{align}
where the subscript `c' is used to indicate that the underlying protocol is channel coding.
A constant composition code with a composition (or the so-called type) $Q$ refers to a codebook whose codewords all have the same empirical distribution $Q$, i.e.~
\begin{equation}
\frac1n \sum_{i=1}^n \mathbf{1}_{\left\{ x = x_i \right\}} = Q(x), \quad \forall x\in\mathcal{X},
\end{equation}
for any indicator function $\mathbf{1}$.
We denote by $\eps_\text{c}^\star\left(n, R , Q\right)$ the {minimum} average probability of error over all $n$-blocklength constant composition codes with type $Q$ and transmission rate $R$, i.e.
\begin{align}
\eps_\text{c}^\star(n,R, Q) := \inf \{ \bar{\eps}_\text{c}(\mathcal{C}_n) : \mathcal{C}_n \text{ is an }(n,R)\text{-code with type $Q$}\}.
\end{align}

In the problems of classical source coding with QSI (or called the classical-quantum Slepian-Wolf coding), the aim is to compress classical data and decompress them with the aid of QSI.
The classical data are represented by $n$-length sequence $\mathbf{x}^n \in \mathcal{X}^{\otimes n}$. The sequence can be produced from an identical and independently distributed (i.i.d.) probability distribution $P \in \mathscr{P}(\mathcal{X})$, i.e.
\begin{align}
\Pr(\mathbf{x}^n) =
\begin{cases}
\Pi_{i=1}^n P(x_i) & \text{sources with i.i.d.~$P$} \\
 \frac{1}{|T_Q^n|} \mathbf{1}_{\{ \mathbf{x}^n \in T_Q^n  \}} & \text{sources with type $Q$} \\
\end{cases},
\end{align}
where we denote by $T_Q^n$ the type class that contains all $n$-length sequences with type $Q$.

The encoder $\mathcal{E}_n:\mathcal{X}^n\to \mathcal{I}$ maps the source to a finite index set $\mathcal{I}$ with \emph{compression rate} $R:= \frac1n \log |\mathcal{I}|$.
The QSI
\begin{align}
\rho_{B^n}^{\mathbf{x}^n} := \rho_B^{x_1} \otimes \rho_B^{x_2} \otimes \cdots \otimes \rho_B^{x_n} \in \mathcal{S}(B)^{\otimes n}
\end{align}
is a product state and can be viewed as a c-q channel applied on the sequence as described in Eq.~\eqref{eq:W^n}.
For the case of i.i.d.~source, the joint distribution that governs the sources and QSI can be modeled as a so-called c-q state $\rho_{XB} := \sum_{x\in\mathcal{X}} P(x) |x\rangle\langle x| \otimes \rho_B^{x} \in \mathcal{S}(XB)$. This is nothing but the joint measure $P\circ \mathscr{W}$ described above when $\mathscr{W}: x\mapsto \rho_B^x \in \mathcal{S}(B)$.
The decoder $\mathcal{D}_n$ is a family of POVM $
\{ \Pi_{\mathbf{x}^n}^{(i)} \}_{i\in\mathcal{I}}$ that receive the index $i \in \mathcal{I}$ and the corresponding density operator $\rho_{B^n}^{\mathbf{x}^n}$ to reproduce the source $\hat{\mathbf{x}}^n$. The error probability of the code $\mathcal{C}_n = (\mathcal{E}_n, \mathcal{D}_n)$ is thus
\begin{align}
{\eps}_\text{s}( \mathcal{C}_n ) :=
\Pr\left( \hat{\mathbf{x}}^n \neq \mathbf{x}^n \right) = \sum_{\mathbf{x}^n\in\mathcal{X}^n} \Pr\left(\mathbf{x}^n\right) \Tr\left[ \rho_{B^n}^{\mathbf{x}^n} \Pi_{\mathbf{x}^n}^{(\mathcal{E}_n(\mathbf{x}^n))} \right].
\end{align}
We denote by $\eps^\star_\text{s}\left(n, R \right)$ the {minimum} average probability of error among all $n$-length source codes and compression rate $R$, i.e.
\begin{align}
\eps_\text{s}^\star (n,R) := \inf \{ \bar{\eps}_\text{s}(\mathcal{C}_n) : \mathcal{C}_n \text{ is an }(n,R)\text{-code with i.i.d.~source $P$}\},
\end{align}
where the subscript `s' is used to indicate that the underlying protocol is source coding.
Similarly, we denote by $\eps_\text{s}^\star\left(n, R , Q\right)$ the {minimum} average probability of error over all $n$-length source codes with type $Q$ and transmission rate $R$, i.e.
\begin{align}
\eps_\text{s}^\star(n,R,Q) := \inf \{ \bar{\eps}_\text{s}(\mathcal{C}_n) : \mathcal{C}_n \text{ is an }(n,R, Q)\text{-code with type $Q$}\}.
\end{align}

\subsection{Error Exponent Analysis with Auxiliary Functions}
One of the fundamental and critical problems in quantum information theory is to characterize the exponent of $\eps_\text{c}$ in terms of the coding blocklength and a fixed rate (We refer the reader to an exposition in Ref.~\cite{Hao-Chung}).
Burnashev and Holevo \cite{BH98, Hol00} first proved a random coding bound for pure-state channels (i.e.~the channel output consists of rank-one density operators):
\begin{align} \label{eq:BH98}
-\frac1n \log \eps_\text{c}^\star(n,R) \geq  \sup_{P\in\mathscr{P}(\mathcal{X})} \sup_{ 0\leq s \leq 1} \left\{ E_{0 }^{\texttt{r}}(s,P)  - sR  \right\}  - \frac1n \log 4, \quad \forall n\in\mathbb{N}.
\end{align}
Here, the auxiliary function $E_{0 }^{\texttt{r}}$ is defined in Eq.~\eqref{eq:E0_c1} with Petz's R\'enyi divergence \cite{Pet86}.
We note that the c-q channel $\mathscr{W}: x \mapsto W_x$ can be viewed as a collection of density operators indexed by $x\in\mathcal{X}$. Hence, the definitions given in Eqs.~\eqref{eq:E0_c1} and \eqref{eq:E0_c2} naturally apply.
The achievability bound in Eq.~\eqref{eq:BH98} was conjectured to hold for general classical-quantum channels. However, it is still open.

For the optimality (i.e.~lower bound of $\eps_\text{c}$), Winter \cite{Win99} employed a dummy channel method by Haroutunian \cite{Har68} to prove a sphere-packing bound for c-q channel:
\begin{align} \label{eq:Win99}
\lim_{n\to \infty} -\frac1n \log \eps_\text{c}^\star(n,R) \leq \sup_{P\in\mathscr{P}(\mathcal{X})}
\inf_{\mathscr{V}:\mathcal{X}\to \mathcal{S}(B) } \left\{ \sum_x P(x) D\left( V_x\|W_x\right) : I(P,\mathscr{V})\leq R \right\}.
\end{align}
We note that the quantity before optimizing for all $P$ equals Csisz\'ar's expression \cite{Csi98} in Eq.~\eqref{eq:information_var}.
Via a variational representation, the right-hand side of Winter's bound can be rewritten in terms of the auxiliary function $E_0^{\texttt{a},\flat}$ defined by the log-Euclidean R\'enyi divergence:
\begin{align}
\sup_{P\in\mathscr{P}(\mathcal{X})} \sup_{ s\geq 1} \left\{ E_{0 }^{\texttt{a},\flat}(s,P)  - sR  \right\}.
\end{align}
Later, Dalai \cite{Dal13} generalized the approach by Shannon, Gallager, and Berlekamp \cite{SGB67, SGB67b} to prove another version of sphere-packing bound for c-q channels:
\begin{align} \label{eq:Dal13}
\lim_{n\to \infty} -\frac1n \log \eps_\text{c}^\star(R) \leq \sup_{P\in\mathscr{P}(\mathcal{X})} \sup_{ s \geq 0} \left\{ E_{0 }^{\texttt{r}}(s,P)  - sR  \right\}.
\end{align}
Part of the authors further refined Dalai's result to the finite blocklength regime with higher-order terms of order $O\left(\frac{\log n}{n}\right)$.

According to Lemma~\ref{lemm:prop}-\ref{prop-f} and Eq.~\eqref{eq:radius}, we have, for all $s\geq 0$,
\begin{align}
\sup_{P\in\mathscr{P}(\mathcal{X})} E_{0 }^{\texttt{r}}(s,P) \leq
\sup_{P\in\mathscr{P}(\mathcal{X})} E_{0 }^{\texttt{a},\flat}(s,P).
\end{align}
Hence, the entropic exponent defined with Petz's version is tighter in the optimality bound. Moreover, {the right-hand side of Eqs.~\eqref{eq:Dal13} and \eqref{eq:BH98} coincides when $R\geq R_\text{crit}$, where the critical rate $R_\text{crit}$ is the rate at which the slope of the right-hand side of \eqref{eq:Dal13}  is $-1$. That is the reason why the entropic exponent with Petz's version is believed to be the optimal error exponent. Let us emphasize again that the achievability bound given in \eqref{eq:BH98} holds only for pure-state channels

If we restrict the channel codes to have a fixed type $P$, it is proved that the entropic exponent function defined in terms of $E_0^{\texttt{a}}$ gives an upper bound to the exponent of $\eps_\text{c}^\star(n,R,P)$ \cite{DW14, CHT17}:
\begin{align}
&-\frac1n \log \eps_\text{c}^\star(n,R,P) \leq \sup_{ s \geq 1} \left\{ E_{0 }^{\texttt{a}}(s,P)  - sR  \right\}  + O\left(\frac{\log}{n}\right),
\end{align}
where the higher-order term can be explicitly determined in \cite{CHT17}. However, it is still open for general codes even in the classical case.

In the strong converse region ($R>C_\mathscr{W}$), the operational strong converse exponent has been determined by Mosonyi and Ogawa \cite{MO17}:
\begin{align}
\lim_{n\to\infty}  -\frac1n \log \left[ 1 - \eps_\text{c}^\star(n,R) \right] &= \inf_{P\in\mathscr{P}(\mathcal{X})} \sup_{ -1 < s < 0} \left\{ E_{0 }^{\texttt{r},*}(s,P)  - sR  \right\} \\
&= \inf_{P\in\mathscr{P}(\mathcal{X})} \sup_{ -1 < s < 0} \left\{ E_{0 }^{\texttt{a},*}(s,P)  - sR  \right\}.
\end{align}
Note that the auxiliary functions in the above equality have been switched to the sandwiched version.

Recently, the finite blocklength sphere-packing bound was proved for the classical source coding with QSI as well \cite{CHDH-2018, CHDH2-2018}.
For rate greater the compression limit, i.e.~$R>H(X|B)_\rho:= -D\left(\rho_{XB}\|\mathds{1}_X\otimes \rho_B\right),$ the bound for sources with a fixed type $Q$ is,
\begin{align}
& -\frac1n \log \eps_\text{s}^\star(n,R,P) \leq \sup_{s\geq 0} \left\{ E_{0,\text{s}}(s,P)  + sR  \right\} + O\left(\frac{\log}{n}\right),
\end{align}
and for i.i.d.~sources is,
\begin{align}
& -\frac1n \log \eps_\text{s}^\star(n,R) \leq \sup_{s\geq 0} \left\{ E_{0,\text{s}}(s)  + sR  \right\} + O\left(\frac{\log}{n}\right).
\end{align}
Here, the auxiliary functions for source coding with QSI are defined as follows \cite{CHDH-2018, CHDH2-2018}:
\begin{align}
E_{0,\text{s}}^{(t)}(s) &:= -sH^{(t)}_{\frac{1}{1+s}}(X|B)_\rho \\
E_{0,\text{s}}^{(t)}(s,P) &:= E_{0,\text{c}}^{\texttt{a},(t)} (s,P) - sH(P) \\
H^{(t)}_\alpha(X|B)_\rho &:= - \inf_{\sigma \in \mathcal{S}(B)} D_\alpha\left( \rho_{XB} \| \mathds{1}_X \otimes \sigma \right).
\end{align}

The operational strong converse exponent (when $R<H(X|B)_\rho$) has been completely determined in terms of the sandwiched version \cite{CHDH-2018}:
\begin{align}
\lim_{n\to\infty} &-\frac1n \log \left[ 1 - \eps_\text{s}^\star(n,R) \right] = \sup_{ -1 < s < 0} \left\{ E_{0,\text{s}}^*(s)  + sR  \right\}.
\end{align}

To ease the burden of notation, we define the following entropic functions for classical-quantum channel coding and classical source coding with QSI, respectively: for $R\geq 0$ and $P\in\mathscr{P}(\mathcal{X})$,
\begin{align}
E_\text{c}(R,P) &:= \sup_{ s > -1} \left\{
E_{0}^{\texttt{a}}(R,P) \vee E_{0}^{\texttt{a},*}(R,P) - s R
\right\}, \label{eq:exponent_c_P}\\
E_\text{c}(R) &:= \begin{dcases}
\sup_{P\in\mathscr{P}(\mathcal{X})} E_\text{c}(R,P), & R \leq C_\mathscr{W} \\
\inf_{P\in\mathscr{P}(\mathcal{X})} E_\text{c}(R,P), & R > C_\mathscr{W}
\end{dcases}, \\
E_\text{s}(R,P) &:= \sup_{ s > -1} \left\{
E_{0,\text{s}}(R,P) \vee E_{0,\text{s}}^{*}(R,P) + s R
\right\}, \\
E_\text{s}(R) &:= \sup_{ s > -1} \left\{
E_{0,\text{s}}(R) \vee E_{0,\text{s}}^{*}(R) + s R
\right\}.
\end{align}

In the following, we collect applications of the properties of the auxiliary functions established in Section~\ref{sec:properties}.
Firstly, in Proposition~\ref{prop:minimax} we prove a minimax identity for the strong converse exponent in classical-quantum channel coding.
\begin{prop}
	[A Minimax Identity in the Strong Converse Regime] \label{prop:minimax}
	For every $R> C_\mathscr{W}$,
	\begin{align}
	E_\textnormal{\text{c}}(R) &= \inf_{P\in\mathscr{P}(\mathcal{X})} \sup_{-1<s<0} \left\{ E_0^{\textnormal{\texttt{r}},*}(s,P) - s R  \right\}
	= \sup_{-1<s<0} \inf_{P\in\mathscr{P}(\mathcal{X})} \left\{ E_0^{\textnormal{\texttt{r}},*}(s,P) - s R  \right\} \\
	&= \inf_{P\in\mathscr{P}(\mathcal{X})} \sup_{-1<s<0} \left\{ E_0^{\textnormal{\texttt{a}},*}(s,P) - s R  \right\}
	= \sup_{-1<s<0} \inf_{P\in\mathscr{P}(\mathcal{X})} \left\{ E_0^{\textnormal{\texttt{a}},*}(s,P) - s R  \right\}.
	\end{align}
\end{prop}
Note that in a recent paper \cite{MO18}, Mosonyi and Ogawa proved an expression for the strong converse exponent of  constant composition with type $P$:
\begin{align}
\lim_{n\to\infty} &-\frac1n \log \left[ 1 - \eps_\text{c}^\star(n,R,P) \right] = E_\text{c}(R,P), \quad R> C_\mathscr{W}.
\end{align}
The above result together with the established Proposition~\ref{prop:minimax} then imply an important consequence. Namely, the strong converse exponent (over all possible codes) can be asymptotically attained by the best constant composition code, i.e.~for $R> C_\mathscr{W}$,
\begin{align}
\lim_{n\to\infty} -\frac1n \log \left[ 1 - \eps_\text{c}^\star(n,R) \right] = 	E_\textnormal{\text{c}}(R) = \inf_{P \in \mathscr{P}(\mathcal{X}) } E_\text{c}(R,P)
= \lim_{n\to\infty} \inf_{ \text{type } Q }  -\frac1n \log \left[ 1 - \eps_\text{c}^\star(n,R,Q) \right].
\end{align}
This gives classical-quantum channel coding with constant composition codes an operational meaning in the strong converse regime.

Secondly, the joint continuity properties provided in Theorem~\ref{theo:joint_cont} and Berge's maximum theorem \cite[Section IV.3]{Ber63}, \cite[Lemma 3.1]{Psh71} show that the entropic exponent functions introduced before are jointly continuous.
Such the joint continuity property is crucial in the variable-length source coding with quantum side information \cite{CHDH2-2018} and finite blocklength analysis.
\begin{prop}[Joint Continuity for Entropic Exponents] \label{prop:jc_exponent}
	Suppose $|\mathcal{H}|<\infty$. The following hold.
	\begin{enumerate}[(a)]
		\item The map $(R,P) \mapsto E_\textnormal{c}(R,P)$ is jointly continuous on $ \left(I_0^{\textnormal{\texttt{a}}}(P,\mathscr{W}),I_\infty^{\textnormal{\texttt{a}},*}(P,\mathscr{W})\right] \times  \mathscr{P}(\mathcal{X})$, and $R\mapsto E_\textnormal{c}(R)$ is continuous on $\left(I_0^{\textnormal{\texttt{a}}}(P,\mathscr{W}),I_\infty^{\textnormal{\texttt{a}},*}(P,\mathscr{W})\right]$.
		
		\item The map $(R,P) \mapsto E_\textnormal{s}(R,P)$ is jointly continuous on $ \left[H(P) - ,I_\infty^{\textnormal{\texttt{a}},*}(P,\mathscr{W}) , H(P) - ,I_0^{\textnormal{\texttt{a}}}(P,\mathscr{W}) \right) \times  \mathscr{P}(\mathcal{X})$, 
		and $R\mapsto E_\textnormal{s}(R)$ is continuous on $\left[H(P) - ,I_\infty^{\textnormal{\texttt{a}},*}(P,\mathscr{W}) , H(P) - ,I_0^{\textnormal{\texttt{a}}}(P,\mathscr{W}) \right)$.
	\end{enumerate}
\end{prop}

Thirdly, Proposition~\ref{prop:source_duality} below shows that the entropic exponent for i.i.d.~source coding with QSI can be reproduced by the type-dependent source; see~\cite{CHDH2-2018}.
\begin{prop}[Entropic Duality in Source Coding with Quantum Side Information \cite{CHDH2-2018}] \label{prop:source_duality}
	Let $\rho_{XB} = \sum_{x\in\mathcal{X}} P_X(x) |x\rangle\langle x| \otimes \rho_B^x \in \mathcal{S}(XB)$ be a joint state of a classical source coding with quantum side information. For any $R\geq 0$, the following holds:
	\begin{align}
	E_\textnormal{s}(R) &= \min_{Q \in \mathscr{P}(\mathcal{X}) } \left\{ E_\textnormal{s}(R, Q) + D(Q\|P_X) \right\}. \label{eq:E_0_Q2_sc}
	\end{align}	
\end{prop}

Lastly, using the concavity of the auxiliary functions given in Section~\ref{sec:properties}, Fenchel's duality theorem \cite{Roc70} directly yields the following useful duality representation in joint source-channel coding with QSI \cite{CHDH3-2018}.
\begin{prop}[Fenchel Duality in Joint Source-Channel Coding with Quantum Side Information \cite{CHDH3-2018}] \label{prop:Fenchel}
	Consider a joint source-channel coding with a classical-quantum joint state $\rho_{XB} \in \mathcal{S}(XB)$ and a classical-quantum channel $\mathscr{W}:\mathcal{X} \to \mathcal{S(H)}$.
	Let $Q\in\mathscr{P}(\mathcal{X})$ and denote by $\sigma_{XB} := \sum_{x\in\mathcal{X}} Q(x) |x\rangle \langle x| \otimes \rho_B^x$.
	Provided $ H(X|B)_\sigma   < I(Q,\mathscr{W})$, we have
	\begin{align}
	\sup_{ s \geq 0} \left\{ E_{0,\textnormal{s}}(s,Q)  + E_{0}^{\texttt{a}}(s,Q)  \right\}
	&= \inf_{ H(X|B)_\sigma < R < I(Q,\mathscr{W})  } \left\{ E_\textnormal{s}(R,Q) + E_\textnormal{c}(R,Q)  \right\}.
	\end{align}
	On the other hand, for $ H(X|B)_\sigma > I(Q,\mathscr{W})$,
	\begin{align}
	\sup_{-1< s < 0} \left\{ E_{0,\textnormal{s}}^{*}(s,Q)  + E_{0}^{\texttt{a},*}(s,Q)  \right\}
	&= \inf_{ I(Q,\mathscr{W}) < R < H(X|B)_\sigma } \left\{ E_\textnormal{s}(R,Q) + E_\textnormal{c}(R,Q)  \right\}.
	\end{align}	
\end{prop}
We remark that Proposition~\ref{prop:Fenchel} is a main ingredient to establish the strong converse exponent in joint source-channel coding with QSI \cite{CHDH3-2018}

\begin{proof}[Proof of Proposition~\ref{prop:Fenchel}]
	We first recall Fenchel's duality theorem \cite{Roc70} named after Werner Fenchel.
	Let $f$ (resp.~$g$) be proper convex function (resp.~proper concave function) from some Banach space to extended real lines.
	Then,
	\begin{align}
	\inf_x \left\{ f(x) - g(x)  \right\} = \sup_{x^*}\left\{ g_*(x^*) - f^*(x^*)   \right\},
	\end{align}
	where
	\begin{align}
	f^*(x^*) := \sup_{x} \left\{ \langle x^*, x \rangle - f(x)  \right\}
	\end{align}
	is the convex conjugate of $f$, and $\langle \cdot, \cdot \rangle$ denotes a inner product. Similarly,
	\begin{align}
	g_*(x^*) := \inf_{x} \left\{ \langle x^*, x \rangle - g(x)  \right\}
	\end{align}
	is the concave conjugate of $g$.
	
	In view of the definitions given above, we let $x=s$, $x^* = R$,
	$-f(x) = E_{0,\text{s}}(s,Q)$ for $s\geq 0$, or $-f(x) = E_{0,\text{s}}^*(s,Q)$ for $s \in (-1,0)$, and let $g(x) = E_{0}^\texttt{a}(s,P)$ for $s\geq 0$, or $ g(x) = E_{0}^{\texttt{a},*}(s,P)$ for $s\in(-1,0)$.
	It can be verified that $f^*(x^*) = E_\text{s}(R,Q)$
	and $-g_*(x^*) =   E_\text{c}(R,Q)$.
	Then, it suffices to show that the auxiliary functions are concave in $s$.
	The concavity of $E_{0,\text{s}}$ and $E_{0}^\texttt{a}$ on $s\geq 0$ follows from \cite[Proposition B.5]{MO17}.
	The concavity of $E_{0,\text{s}}^*$ and $E_{0}^{\texttt{a},*}$ on $s \in (-1,0)$ follows from Theorem~\ref{theo:concave_*}, which completes the proof.
	
\end{proof}

\section{Conclusions and Open Problems} \label{sec:conclusions}
We study the R\'enyi information and Augustin information defined via the Petz, sandwiched, and the log-Euclidean R\'enyi divergences. The uniform equicontinuity and the convexity/concavity in prior probability distributions were proved.
For various quantum auxiliary functions, we established the joint continuity in order and prior.
Moreover, we solve the open problems of the concavity in the region of $s\in (-1,0)$ by employing the complex interpolation theory.
The established properties allow us to better understand the entropic exponents of c-q channel coding, classical data compression with quantum side information, and joint source-channel coding with quantum side information. Applications include a minimax identity in the strong converse region, an entropic duality, and a Fenchel duality relation.
We list the following open problems in the case of infinite-dimensional Hilbert spaces.

\begin{itemize}
	\item Does the Augustin mean uniquely exist for the Petz and the sandwiched form?
	
	\item Are $I_\alpha^{\texttt{r},(t)}$ and $I_\alpha^{\texttt{a},(t)}$ continuous at $\alpha =1$ from below for $(t) = \{\,\}$, $\flat$, or $*$? (Note that we only know that  $\alpha \mapsto I_\alpha^{\texttt{r}}$ is analytical due to the noncommutative Sibson's identity given in Proposition~\ref{prop:Sibson}.)
	
	\item Are $I_\alpha^{\texttt{r},(t)}$ and $I_\alpha^{\texttt{a},(t)}$ continuously differentiable at $\alpha =1$ for $(t) = \{\,\}$ $\flat$ or $*$?
	For example, as proven in the commuting case by Nakibo\u{g}lu \cite[Lemma 17]{Nak18a}., we conjecture that
	\begin{align}
	\frac{\partial}{\partial \alpha } I_\alpha^{\texttt{a}, (t)} (P,\mathscr{W})|_{\alpha = z} = \frac{\partial}{\partial \alpha} D_\alpha^{(t)}\left( \left.\mathscr{W} \right\| q_{z,p}^{(t)} | P \right),
	\end{align}
	where $q_{z,p}^{(t)}$ is the associated Augustin mean.
	
	\item Are $E_0^{\texttt{a}}(s,P)$ and $E_0^{\texttt{a},*}(s,P)$ continuous for $s\in(-1,0)$.
\end{itemize}
Lastly, we remark that although the Augustin information and sandwiched R\'enyi information do not have closed-form expressions, they can be numerically computed in finite-dimensional Hilbert spaces by existing algorithms; see e.g.~\cite{YCL21}.

\section*{Acknowledgments}
H.-C.~Cheng is supported by the Young Scholar Fellowship (Einstein Program) of the Ministry of Science and Technology (MOST) in Taiwan under grant number MOST 109-2636-E-002-001 and 110-2636-E-002-009,
and is supported by the Yushan Young Scholar Program of the Ministry of Education (MOE) in Taiwan under grant number NTU-109V0904 and NTU-110V0904.
H.-C.~Cheng would like to thank the Center for Quantum Science and Engineering, National Taiwan University, and Physics Division, National Center for Theoretical Sciences, Taiwan (R.O.C.) for their supports.
We would like to thank Mil{\'{a}}n Mosonyi and Tomohiro Ogawa for introducing us to the problem and the detailed discussions. We also thank Marius Junge, Nilanjana Datta, Eric P.~Hanson, and Gilles Pisier for the insightful discussions.
We thank Masahito Hayashi for pointing out his work \cite{HT14} for the approach of using the pinching argument.
Lastly, we especially thank Bar\i\c{s} Nakibo\u{g}lu for inspiring us how H\"older's inequality comes into the play in the proof of the concavity of the auxiliary functions, for reading the early version of the manuscript, and for providing useful comments.
We also thank two anonymous reviewers and the editor for carefully reviewing our paper.

\appendix

\section{Proofs of Properties of R\'enyi Information and Augustin Information} \label{sec:proofs}

\begin{prop2}[Properties of R\'enyi Information]
	Let $\mathscr{W} \subset \mathcal{S(H)}$ , and let $(t)$ be any of the three values: $\{\}$, $*$, or $\flat$.
	\begin{enumerate}[(a)]
		\item\label{I2_R-aa} For any $P\in\mathscr{P(W)}$, $I_\alpha^{\textnormal{\texttt{r}},(t)}(P,\mathscr{W})$ is non-negative and nondecreasing in $\alpha$. Moreover, $I_\alpha^{\texttt{r},(t)}(P,\mathscr{W}) \leq\log |\textnormal{\textsf{supp}}(P)|]$.
		
		
		\item\label{I2_R-bb} The map $P\mapsto I_{\alpha}^{\textnormal{\texttt{r}},(t)} (P,\mathscr{W})$ is quasi-concave on $\mathscr{P(W)}$ for $\alpha \in [0,1)$, and concave on $\mathscr{P(W)}$ for $\alpha \in [1,\infty]$.
		
		\item\label{I2_R-cc}
		Let
		\begin{align} \label{eq:set_AA}
		\mathcal{A} := [0,1],\quad \mathcal{A}^* := [1/2, \infty], \; \text{ and } \; \mathcal{A}^\flat := [0,\infty].
		\end{align}		
		For $\eta \in [0,\infty]$, if $C_{\eta, \mathscr{W}}^{(t)} < \infty$, then $\left\{ I_\alpha^{\textnormal{\texttt{r}},(t)}(P,\mathscr{W})  \right\}_{ \alpha\in[0,\eta] \cap \mathcal{A}^{(t)} }$ is uniformly equicontinuous in $P \in\mathscr{P(W)}$.
	\end{enumerate}
\end{prop2}


\begin{proof}[Proof of Proposition~\ref{prop:I2_R}-\ref{I2_R-aa}]
	The assertion about the monotonicity follows directly from Lemma~\ref{lemm:prop}-\ref{prop-a} and the definition of $I_\alpha^{\texttt{r},(t)}$ given in Eq.~\eqref{eq:I2_R}, which was also pointed out by Mosonyi and Ogawa \cite[Lemma 4.6]{MO17}.
	
	We move on to prove the upper bound. Recall Eq.~\eqref{eq:temp}, it suffices to prove it for Petz's version.
	By quantum Sibson's identity \cite{SW12}, it holds for every $\alpha \in (1,\infty)$,
	\begin{align}
	I_\alpha^{\texttt{r}}(P,\mathscr{W}) &= \frac{\alpha}{\alpha-1} \log \Tr\left[ \left( \sum_{\omega \in \mathscr{W} } P(\omega) \omega^\alpha \right)^{\frac{1}{\alpha}}\right] \\
	&\leq  \log \Tr\left[ \left( \sum_{\omega \in \mathscr{W} } P(\omega ) \omega^\alpha \right)^{\frac{1}{\alpha}} \right].	
	\end{align}
	Next, we employ a generalized Ando-Zhan theorem proved by Bhatia and Kittaneh \cite{AZ99}, \cite[Theorem 5]{BK04}: for any positive operators $A_1,\ldots A_m$ and every unitarily invariant norm $\opnorm{\cdot}$,
	\begin{align}
	\vertiii{ \left( \sum_{i=1}^m A_i \right)^r   }
	\leq \vertiii{   \sum_{i=1}^m A_i^r }, \quad \forall r\in[0,1].
	\end{align}
	Therefore, we have
	\begin{align}
	I_\alpha^{\texttt{r}}(P,\mathscr{W})
	&\leq \log \Tr\left[ \sum_{\omega \in \mathscr{W}} P(\omega)^{\frac{1}{\alpha} } \omega \right] \\
	&\leq \log |\textnormal{\textsf{supp}}(P)|,
	\end{align}
	which completes the proof.
\end{proof}

\begin{proof}[Proof of Proposition~\ref{prop:I2_R}-\ref{I2_R-bb}]
	The arguments follow similar from \cite[Theorems 7, 8]{HV15}.
	Note that for every $\sigma \in \mathcal{S(H)}$,
	\begin{align}
	D_\alpha\left( P\circ \mathscr{W} \| P\otimes \sigma \right) = \frac{1}{\alpha-1} \log \sum_{\omega \in \mathscr{W}} P(\omega) \mathrm{e}^{(\alpha-1)D_\alpha^{(t)}(\omega\|\sigma)}.
	\end{align}
	Since $\frac{1}{\alpha}\log \left(\al \right)$ is a decreasing function for $\alpha\in[0,1)$ and concave function for $\alpha\geq 1$, the assertions follow because pointwise infimum of quasiconcave functions is concave.
\end{proof}

\begin{proof}[Proof of Proposition~\ref{prop:I2_R}-\ref{I2_R-cc}]
	Let $t$ be any of three values, and fix any $\alpha \in \mathcal{A}^{(t)}$.
	We prove our claim by showing for any $P_1, P_2 \in \mathscr{P}(\mathcal{X})$,
	\begin{align}
	&\sup_{\alpha\in[0,\eta] \cap \mathcal{A}^{(t)} } \left| I_\alpha^{\texttt{r},(t)}(P_2, \mathscr{W}) - I_\alpha^{\texttt{r},(t)}(P_1, \mathscr{W}) \right| \notag \\
	&\leq
	\begin{cases} \label{eq:R1_0}
	\log \left[ \frac{1}{1-\delta} \wedge \frac{\mathrm{e}^{ C_{0,\mathscr{W}}^{(t)} }}{\delta}   \right]
	+ \log \left[ 1-\delta + \delta \mathrm{e}^{C_{0,\mathscr{W}}^{(t)} } \right], & \eta = 0\\
	\log \frac{ 1-\delta + \delta \mathrm{e}^{C_{\eta,\mathscr{W}}^{(t)} } }{ \left[ (1-\delta)^{\frac{1}{\eta}} + \delta^{\frac{1}{\eta}} \mathrm{e}^{ \frac{\eta-1}{\eta} C_{\eta,\mathscr{W}}^{(t)} }  \right]^{\frac{\eta}{1-\eta}}   }, &\eta \in \mathbb{R}_{>0}\backslash{1} \\
	h(\delta) + \delta C_{1,\mathscr{W}} + \log \left[ 1-\delta + \delta \mathrm{e}^{C_{1,\mathscr{W}}} \right], & \eta = 1
	\end{cases}
	\end{align}
	where
	$\delta := \frac{1}{2}\left\| P_1 - P_2 \right\|_1$ and $h(\delta) := -\delta\log \delta - (1-\delta) \log (1-\delta)$.
	We remark that the presentation follows from the commuting case in Ref.~\cite{Nak16a} for readers' convenience.
	
	Invoke the decomposition provided in \cite[Lemma 4-(c)]{Nak16a}, i.e.
	\begin{align}
	\begin{split} \label{eq:decomp}
	P_1 &= (1-\delta) s_{\wedge} + \delta s_1, \\
	P_2 &= (1-\delta) s_{\wedge} + \delta s_2,
	\end{split}
	\end{align}
	where
	$s_{\wedge} := \frac{ P_1 \wedge P_2 }{ 1 - \delta }$,
	$s_1 := \frac{ P_1 - P_1 \wedge P_2}{ \delta }$,
	$s_2 := \frac{ P_2 - P_1 \wedge P_2}{ \delta }$.
	For notational convenience, we let $Q_\alpha^{(t)}\left( \rho\|\sigma \right) := \mathrm{e}^{(\alpha-1) D_\alpha^{(t)}(\rho\|\sigma)}$.
	Further, we denote the \emph{R\'enyi mean} for any distribution $P\in\mathscr{P}(\mathcal{X})$ and $\alpha>0$ by
\begin{align}
\sigma_{\alpha,P} \in \argmin_{\sigma \in \mathcal{S(H)}} D_\alpha^{(t)}\left( P\circ \mathscr{W} \| P\otimes \sigma \right).
\end{align}

We begin the proof by showing a lower bound on $I_\alpha^{\texttt{r},(t)}\left(P_1, \mathscr{W} \right)$.
Note that the order-$1$ R\'enyi mean $\sigma_{1,P}$ equals to the average state $P\mathscr{W}$ (see e.g.~\cite{SW97}).
Direct calculation shows that
\begin{align}
I_1(P_1,\mathscr{W}) &= (1-\delta) I_1(s_\wedge,\mathscr{W}) + (1-\delta) D_1\left( \sigma_{1,s_\wedge} \| \sigma_{1, P_1} \right) + \delta I_1(s_1,\mathscr{W}) + \delta D_1\left( \sigma_{1,s_1} \| \sigma_{1, P_1} \right) \\
&\geq  (1-\delta) I_1(s_\wedge,\mathscr{W}) + \delta I_1(s_1,\mathscr{W}), \label{eq:R1_2}
\end{align}
where the inequality follows from the non-negativity of the R\'enyi divergence given in Lemma~\ref{lemm:prop}--\ref{prop-b}.

Using Eq.~\eqref{eq:decomp}, it follows that for $\alpha\neq 1$,
\begin{align}
I_\alpha^{\texttt{r},(t)}\left(P_1, \mathscr{W} \right) &= \frac{1}{\alpha-1} \log \left[ (1-\delta) \sum_{\omega} s_{\wedge}(\omega) Q_\alpha^{(t)}\left(\omega\| \sigma_{\alpha, P_1} \right)  + \delta \sum_{\omega} s_{1}(\omega) Q_\alpha^{(t)}\left(\omega\| \sigma_{\alpha, P_1} \right) \right] \\
&\geq \frac{1}{\alpha-1} \log \left[ (1-\delta) \sum_{\omega} s_{\wedge}(\omega) Q_\alpha^{(t)}\left(\omega\| \sigma_{\alpha, s_\wedge} \right)  + \delta \sum_{x} s_{1}(\omega) Q_\alpha^{(t)}\left(\omega\| \sigma_{\alpha, s_1} \right) \right], \label{eq:R1_1}
\end{align}
where the inequality follows from the definition of $I_\alpha^{\texttt{r},(t)}$ given in Eq.~\eqref{eq:I2_R}, i.e.~
\begin{align}
I_\alpha^{\texttt{r},(t)}\left( s_\wedge, \mathscr{W} \right) &= \frac{1}{\alpha-1}\log \sum_\omega s_\wedge(\omega) Q_\alpha^{(t)}(\omega\| \sigma_{\alpha, s_\wedge} ) \leq \frac{1}{\alpha-1}\log \sum_\omega s_\wedge(\omega) Q_\alpha^{(t)} (\omega \| \sigma_{\alpha, P_1} ), \\
I_\alpha^{\texttt{r},(t)}\left( s_1, \mathscr{W} \right) &= \frac{1}{\alpha-1}\log \sum_\omega s_1(\omega) Q_\alpha^{(t)}( \omega \| \sigma_{\alpha, s_1} ) \leq \frac{1}{\alpha-1}\log \sum_\omega s_1(\omega) Q_\alpha^{(t)}(\omega \| \sigma_{\alpha, P_1} ).
\end{align}
Then, from Eqs.~\eqref{eq:R1_2} and \eqref{eq:R1_1}, we have
\begin{align}
I_\alpha^{\texttt{r},(t)}\left(P_1, \mathscr{W} \right) &\geq
\begin{dcases}
(1-\delta) I_1\left(s_\wedge, \mathscr{W} \right) + \delta I_1\left( s_1, \mathscr{W} \right), & \alpha = 1 \\
\frac{1}{\alpha-1} \log \left[ (1-\delta) \mathrm{e}^{ (\alpha-1) I_\alpha^{\texttt{r},(t)}\left( s_\wedge, \mathscr{W} \right)}  + \delta \mathrm{e}^{ (\alpha-1) I_\alpha^{\texttt{r},(t)}\left( s_1, \mathscr{W} \right)} \right], & \alpha \neq 1
\end{dcases} \\
&= I_\alpha^{\texttt{r},(t)}\left(s_\wedge, \mathscr{W} \right) - g\left( \delta, \alpha, I_\alpha^{\texttt{r},(t)}\left(s_\wedge, \mathscr{W} \right) - I_\alpha^{\texttt{r},(t)}\left(s_1, \mathscr{W} \right) \right),
\end{align}
where for any $\delta \in [0,1]$, $\alpha > 0$, $\gamma \in \mathbb{R}$, we define the function $g(\delta,\alpha,\gamma)$ by
\begin{align}
g(\delta,\alpha,\gamma) :=
\begin{cases}
\delta \gamma, & \alpha = 1\\
\frac{1}{1-\alpha} \log \left[ (1-\delta) + \delta \mathrm{e}^{(1-\alpha)\gamma} \right], &\alpha \neq 1
\end{cases}.
\end{align}
Since $\alpha \mapsto g(\delta, \alpha, \gamma)$ is nonincreasing (see \cite[p.~25]{Nak16a}), we have
\begin{align}
I_\alpha^{\texttt{r},(t)}\left(P_1, \mathscr{W} \right) &\geq I_\alpha^{\texttt{r},(t)}\left(s_\wedge, \mathscr{W} \right) - g\left( \delta, 0, I_\alpha^{\texttt{r},(t)}\left(s_\wedge, \mathscr{W} \right) - I_\alpha^{\texttt{r},(t)}\left(s_1, \mathscr{W} \right) \right).
\end{align}
Moreover, the map $\gamma \mapsto g(\delta, \alpha, \gamma)$ is nondecreasing. Hence, by using $I_\alpha^{\texttt{r},(t)}(s_1,\mathscr{W}) \geq 0$, $I_\alpha^{\texttt{r},(t)}(s_\wedge,\mathscr{W}) \leq I_\eta^{\texttt{r},(t)}(s_\wedge,\mathscr{W})$, and $I_\eta^{\texttt{r},(t)}(s_1,\mathscr{W}) \leq C_{\eta, \mathscr{W}}^{(t)}$, we obtain
\begin{align} \label{eq:R1_3}
I_\alpha^{\texttt{r},(t)}\left(P_1, \mathscr{W} \right) &\geq I_\alpha^{\texttt{r},(t)}\left(s_\wedge, \mathscr{W} \right) - g\left( \delta, 0, C_{\eta, \mathscr{W}}^{(t)} \right).
\end{align}

Next, we move on to show an upper bound on $I_\alpha^{\texttt{r},(t)}\left(P_2, \mathscr{W} \right)$.
For $\alpha=1$, we have
\begin{align}
I_1(P_2,\mathscr{W}) &= \inf_{\sigma \in \mathcal{S(H)}} \sum_{\omega\in\mathscr{W}} P(\omega) D(\omega\|\sigma) \\
&\leq \sum_{\omega\in\mathscr{W}} P(\omega) D\left(\omega\|  (1-\delta) \sigma_{1,s_\wedge} + \delta \sigma_{1,s_2} \right) \\
&= (1-\delta) I_1(s_\wedge, \mathscr{W}) + \delta I_1(s_2,\mathscr{W}) + h(\delta). \label{eq:R1_6}
\end{align}

On the other hand, from the definition of $I_\alpha^{\texttt{r},(t)}$ given in Eq.~\eqref{eq:I2_R} and using Eq.~\eqref{eq:decomp} again, we have, for any $\alpha\neq 1$ and $\sigma \in \mathcal{S(H)}$,
\begin{align}
I_\alpha^{\texttt{r},(t)}\left(P_2, \mathscr{W} \right)
&\leq D_\alpha^{(t)}\left( P_2 \circ \mathscr{W} \| P_2 \otimes \sigma \right) \\
&=\frac{1}{\alpha-1} \log \left[ (1-\delta) \sum_x s_\wedge(x) Q_\alpha^{(t)}(\omega \| \sigma ) + \delta \sum_x s_2(x) Q_\alpha^{(t)}(\omega \| \sigma ) \right]. \label{eq:R1_5}
\end{align}
Here, we choose
\begin{align}
\sigma = \frac{  \theta \sigma_{\alpha,s_\wedge} + \vartheta \sigma_{\alpha,s_2} }{ \theta + \vartheta } \in \mathcal{S(H)},\
\theta := (1-\delta)^{\frac{1}{\alpha}} \mathrm{e}^{\frac{\alpha-1}{\alpha} I_\alpha^{\texttt{r},(t)}(s_\wedge, \mathscr{W}) },\
\vartheta := \delta^{\frac{1}{\alpha}} \mathrm{e}^{\frac{\alpha-1}{\alpha} I_\alpha^{\texttt{r},(t)}(s_2, \mathscr{W}) }.
\end{align}
Note that $\theta,\vartheta \geq 0$. Hence,
\begin{align}
\begin{split} \label{eq:R1_4}
\sigma \geq \frac{\theta}{\theta + \vartheta} \sigma_{\alpha,s_\wedge}, \;\text{ and }\;
\sigma \geq \frac{\vartheta}{\theta + \vartheta} \sigma_{\alpha,s_2}.
\end{split}
\end{align}
Lemma~\ref{lemm:prop}-\ref{prop-c} implies that $Q_\alpha^{(t)}$ is nonincreasing in its second argument for $\alpha > 1$ and nondecreasing in its second argument for $\alpha<1$.
Using this fact and combining Eqs.~\eqref{eq:R1_6}, \eqref{eq:R1_5} and \eqref{eq:R1_4}, direct calculation yields
\begin{align}
I_\alpha^{\texttt{r},(t)}\left(P_2, \mathscr{W} \right)
&\leq
\begin{cases}
(1-\delta) I_1(s_\wedge, \mathscr{W}) + \delta I_1(s_2,\mathscr{W}) + h(\delta), & \alpha = 1 \\
\frac{\alpha}{\alpha-1} \log \left[ \theta + \vartheta\right], & \alpha \neq 1
\end{cases} \\
&= I_\alpha^{\texttt{r},(t)}\left(s_\wedge, \mathscr{W} \right) + f\left( \delta, \alpha, I_\alpha^{\texttt{r},(t)}\left(s_2, \mathscr{W} \right) - I_\alpha^{\texttt{r},(t)}\left(s_\wedge, \mathscr{W} \right) \right),
\end{align}
where for any $\delta \in [0,1]$, $\alpha > 0$, $\gamma \in \mathbb{R}$, we define the function $f(\delta,\alpha,\gamma)$ by
\begin{align}
f(\delta,\alpha,\gamma) :=
\begin{cases}
\delta \gamma + h(\delta), & \alpha = 1 \\
\frac{\alpha}{\alpha-1} \log \left[ (1-\delta)^{\frac{1}{\alpha}} + \delta^{\frac{1}{\alpha}} \mathrm{e}^{\frac{\alpha-1}{\alpha} \gamma} \right], &\alpha\neq 1
\end{cases}.
\end{align}
Since $\alpha \mapsto f(\delta, \alpha, \gamma)$ is nondcreasing (see \cite[p.~26]{Nak16a}), we have
\begin{align}
I_\alpha^{\texttt{r},(t)}\left(P_2, \mathscr{W} \right) &\leq I_\alpha^{\texttt{r},(t)}\left(s_\wedge, \mathscr{W} \right) + f\left( \delta, \eta, I_\alpha^{\texttt{r},(t)}\left(s_2, \mathscr{W} \right) - I_\alpha^{\texttt{r},(t)}\left(s_\wedge, \mathscr{W} \right) \right).
\end{align}
Further, the map $\gamma \mapsto g(\delta, \alpha, \gamma)$ is nondecreasing. Hence, by using $I_\alpha^{\texttt{r},(t)}(s_\wedge,\mathscr{W}) \geq 0$, $I_\alpha^{\texttt{r},(t)}(s_2,\mathscr{W}) \leq I_\eta^{\texttt{r},(t)}(s_2,\mathscr{W})$, and $I_\eta^{\texttt{r},(t)}(s_2,\mathscr{W}) \leq C_{\eta, \mathscr{W}}^{(t)}$, we obtain
\begin{align} \label{eq:R1_7}
I_\alpha^{\texttt{r},(t)}\left(P_2, \mathscr{W} \right) &\leq I_\alpha^{\texttt{r},(t)}\left(s_\wedge, \mathscr{W} \right) + f\left( \delta, \eta, C_{\eta, \mathscr{W}}^{(t)} \right).
\end{align}
Combining Eqs.~\eqref{eq:R1_3} and \eqref{eq:R1_7} gives
\begin{align}
I_\alpha^{\texttt{r},(t)}\left(P_2, \mathscr{W} \right) - I_\alpha^{\texttt{r},(t)}\left(P_1, \mathscr{W} \right) \leq f\left(\delta,\eta, C_{\eta,\mathscr{W}}^{(t)}\right) + g\left(\delta,0, C_{\eta,\mathscr{W}}^{(t)}\right).
\end{align}
A lower bound on $I_\alpha^{\texttt{r},(t)}\left(P_2, \mathscr{W} \right) - I_\alpha^{\texttt{r},(t)}\left(P_1, \mathscr{W} \right)$ can be shown by using a similar argument and reversing the roles of $P_1$ and $P_2$.

It remains to show Eq.~\eqref{eq:R1_0} for $\alpha=0$.
We remark that the argument of this case follows from the similar ideas in \cite[Lemma 16-(e)]{Nak16a}. We provide the proof here for completeness.
From the definition given in Eq.~\eqref{eq:I2_R}, we have
\begin{align}
I_0^{\texttt{r},(t)}(P_1, \mathscr{W}) &= - \sup_{\sigma \in \mathcal{S(H)}}
\log \left[ (1-\delta) \sum_x s_\wedge(x) Q_0^{(t)}(\omega\|\sigma) + \delta \sum_x s_1(x) Q_0^{(t)}(\omega\|\sigma)    \right] \\
&\geq -
\log \left[ (1-\delta) \sup_{\sigma \in \mathcal{S(H)}} \sum_\omega s_\wedge(\omega) Q_0^{(t)}(\omega\|\sigma) + \delta \sup_{\sigma \in \mathcal{S(H)}} \sum_\omega s_1(\omega) Q_0^{(t)}(\omega\|\sigma)    \right] \\
&= - \log \left[ (1-\delta)  \mathrm{e}^{- I_0^{\texttt{r},(t)}(s_\wedge,\mathscr{W})} + \delta \mathrm{e}^{- I_0^{\texttt{r},(t)}(s_1,\mathscr{W})} \right]  \\
&\geq - \log \left[ (1-\delta)  \mathrm{e}^{- I_0^{\texttt{r},(t)}(s_\wedge,\mathscr{W})} + \delta  \right] \\
&= I_0^{\texttt{r},(t)}(s_\wedge,\mathscr{W}) - \log \left[ 1 - \delta + \delta \mathrm{e}^{I_0^{\texttt{r},(t)}(s_\wedge,\mathscr{W}) } \right], \label{eq:R1_8}
\end{align}
where the inequalities follows from the subadditivity of supremum and $I_0^{\texttt{r},(t)}(s_1,\mathscr{W}) \geq 0$.

On the other hands,
\begin{align}
I_0^{\texttt{r},(t)}(P_1, \mathscr{W}) &= \inf_{\sigma \in \mathcal{S(H)}} \log \frac{1}{  (1-\delta)  \sum_\omega s_\wedge(\omega) Q_0^{(t)}( \omega \|\sigma) + \delta  \sum_\omega s_2(x) Q_0^{(t)}( \omega \|\sigma)  } \\
&\leq \left( \inf_{\sigma \in \mathcal{S(H)}} \log \frac{1}{  (1-\delta)  \sum_\omega s_\wedge(\omega) Q_0^{(t)}(\omega\|\sigma)   }  \right) \wedge
	\left( \inf_{\sigma \in \mathcal{S(H)}}  \log \frac{1}{  (1-\delta)  \sum_\omega s_2(\omega) Q_0^{(t)}(\omega\|\sigma)   }  \right) \\
&= \left( I_0^{\texttt{r},(t)}(s_\wedge, \mathscr{W}) + \log\frac{1}{1-\delta} \right) \wedge \left( I_0^{\texttt{r},(t)}(s_2, \mathscr{W}) + \log\frac{1}{\delta} \right). \label{eq:R1_9}
\end{align}
Hence, Eqs.~\eqref{eq:R1_8} along with \eqref{eq:R1_9} lead to
\begin{align}
I_0^{(t)}(P_2, \mathscr{W}) - I_0^{(t)}(P_1, \mathscr{W}) \leq \log \left[ 1-\delta + \delta \mathrm{e}^{ C_{\eta, \mathscr{W}}^{(t)}} \right] + \log \left[ \frac{1}{1-\delta} \wedge \frac{ \mathrm{e}^{C_{\eta,\mathscr{W}}^{(t)} } }{\delta} \right].
\end{align}
A lower bound on $I_0^{\texttt{r},(t)}\left(P_2, \mathscr{W} \right) - I_0^{\texttt{r},(t)}\left(P_1, \mathscr{W} \right)$ can be shown by using a similar argument and reversing the roles of $P_1$ and $P_2$. As a result, Eq.~\eqref{eq:R1_0} holds for $\eta = 0 $ and $\alpha = 0$.

Finally, the case of $\eta > 0$ and $\alpha = 0$ follows by noting that
\begin{align}
\frac{\eta}{\eta - 1} \log \left[ (1-\delta)^{\frac{1}{\eta}} + \delta^{\frac{1}{\eta}} \mathrm{e}^{\frac{\eta - 1}{\eta} C_{\eta, \mathscr{W}} } \right] \geq \log \left[ \frac{1}{1-\delta} \wedge \frac{ \mathrm{e}^{C_{\eta,\mathscr{W}}^{(t)}  } }{\delta} \right].
\end{align}
\end{proof}

\begin{prop3}[Properties of Augustin Information]
Let $\mathscr{W}\subset \mathcal{S(H)}$ be a classical-quantum channel, $(t)$ be any of the three values: $\{\}$, $*$, or $\flat$, and let $\mathcal{A}^{(t)}$ be defined in \eqref{eq:set_A}.
\begin{enumerate}[(a)]
	\item\label{I2-aa} For every $P\in\mathscr{P}(\mathcal{X})$,
	$I_\alpha^{\textnormal{\texttt{a}},(t)}(P,\mathscr{W})$ is non-negative and nondecreasing in $\alpha$.
	Moreover, $I_\alpha^{\textnormal{\texttt{a}},(t)}(P,\mathscr{W}) \leq H(P)$ for $\alpha \in \mathcal{A}^{(t)}$, where $H(P)$ is the Shannon entropy of $P$.
		
		
		\item\label{I2-bb} For any $\alpha> 0$, the map $P \mapsto I_{\alpha}^{\textnormal{\texttt{a}},(t)} (P,\mathscr{W})$ is concave on $\mathscr{P(W)}$.
		
\item\label{I2-cc} Let $\mathcal{A}^{(t)}$ be defined in \eqref{eq:set_AA}. For $\eta \in [0,\infty]$, if $C_{\eta, \mathscr{W}}^{(t)} < \infty$, then $\left\{ I_\alpha^{\texttt{a},(t)}(P,\mathscr{W})  \right\}_{ \alpha\in (0,\eta] \cap \mathcal{A}^{(t)} }$ is uniformly equicontinuous in $P\in\mathscr{P(W)}$.
	\end{enumerate}
\end{prop3}

\begin{proof}[Proof of Proposition~\ref{prop:I2}-\ref{I2-aa}]
	As in Proposition~\ref{prop:I2_R}-\ref{I2_R-aa}, the assertion about the monotonicity follows direct from Lemma~\ref{prop-a} and the definition of $I_\alpha^{\texttt{a},(t)}$ given in Eq.~\eqref{eq:I2}.
	
	The upper bound follows similar idea as in \cite[Lemma 13]{Nak18a}.
	The definition of $I_\alpha^{\texttt{a},(t)}$ implies that
	\begin{align}
	I_\alpha^{\texttt{a},(t)}(P,\mathscr{W}) &= \inf_{\sigma \in\mathcal{S(H)}} \sum_{\omega} P(\omega) D_\alpha^{(t)} \left( \omega \|\sigma \right) \\
	&\leq  \sum_{\omega} P(\omega) D_\alpha^{(t)} \left( \omega \| P\mathscr{W} \right) \\
	&\leq \sum_{\omega} P(\omega) \log \frac{1}{P(\omega)},
	\end{align}
	where the last inequality follows from {Lemma~\ref{lemm:prop}-\ref{prop-c}}.
\end{proof}
\begin{proof}[Proof of Proposition~\ref{prop:I2}-\ref{I2-bb}]
	The concavity immediately follows from the definition of $I_\alpha^{\texttt{a},(t)}$ given in Eq.~\eqref{eq:I2}, and the fact that pointwise infimum of concave functions is concave.
\end{proof}

\begin{proof}[Proof of Proposition~\ref{prop:I2}-\ref{I2-cc}]
Fix $t$ be any of the three values.
To prove the equicontinuity, we need the following inequality:
\begin{align}
I_\alpha^{\texttt{a},(t)}(\alpha,P_\beta) &\leq \beta I_\alpha^{\texttt{a},(t)}(\alpha,P_1) + (1-\beta) I_\alpha^{\texttt{a},(t)}(\alpha,P_0) + H(\beta) \label{eq:ec2}
\end{align}
for any $P_1, P_0 \in \mathscr{P}(\mathcal{X})$, $P_\beta = \beta P_1 + (1-\beta) P_0$, $\beta \in (0,1)$, $\alpha\in [0,1]$; and we shorthand $H(\beta):= -\beta \log \beta - (1-\beta)\log (1-\beta)$ the binary entropy function.

We denote the \emph{Augustin mean} for any distribution $P\in\mathscr{P}(\mathcal{X})$ and $\alpha>0$ by
\begin{align}
\sigma_{\alpha,P} \in \argmin_{\sigma \in \mathcal{S(H)}} \sum_{\omega\in\mathscr{W}} P(\omega) D_\alpha^{(t)}\left( \omega \| \sigma \right).
\end{align}
Lemma~\ref{lemm:prop}-\ref{prop-b} implies that, for every $\alpha \in [0,1]$,
\begin{align}
&\sum_{\omega\in\mathscr{W}} P_\beta(\omega) D_{\alpha}^{(t)}\left(\omega\| \beta \sigma_{\alpha,P_1} + (1-\beta) \sigma_{\alpha,P_0} \right) \notag \\
&= \beta \sum_{ \omega \in \mathscr{W} } P_1(\omega) D_{\alpha}^{(t)}\left(\omega\| \beta \sigma_{\alpha,P_1}  + (1-\beta) \sigma_{\alpha,P_0} \right) + (1-\beta) \sum_{ \omega \in\mathscr{W} } P_0(\omega) D_{\alpha}^{(t)}\left( \omega\| \beta \sigma_{\alpha,P_1} + (1-\beta) \sigma_{\alpha,P_0} \right) \\
&\leq \beta \sum_{\omega \in \mathscr{W} } P_1(\omega) D_{\alpha}^{(t)} \left( \omega\| \sigma_{\alpha, P_1} \right) - \beta\log\beta + (1-\beta) \sum_{ \omega\in\mathscr{W} } P_0(\omega) D_{\alpha}^{(t)}\left( \omega\| \sigma_{\alpha, P_0} \right) - (1-\beta)\log (1-\beta) \\
&= \beta I_\alpha^{\texttt{a}, (t)}(P_1,\mathscr{W}) + (1-\beta) I_\alpha(P_0, \mathscr{W}) + H(\beta).
\end{align}

Let $s_\wedge, s_1, s_0$ be
\begin{align}
s_\wedge &= \frac{ P_1\wedge P_0 }{ \left\| P_1\wedge P_0 \right\|_1}, \\
s_1 &= \frac{ P_1 - P_1\wedge P_0 }{ 1 - \left\| P_1\wedge P_0 \right\|_1 }, \\
s_0 &= \frac{ P_0 - P_1\wedge P_0 }{ 1 - \left\| P_1\wedge P_0 \right\|_1 }.
\end{align}
One can verify that.
\begin{align}
P_1 &= \left( 1 - \frac{\left\| P_1- P_0 \right\|_1}{2} \right) s_\wedge + \frac{\left\| P_1- P_0 \right\|_1}{2} s_1, \\
P_0 &= \left( 1 - \frac{\left\| P_1- P_0 \right\|_1}{2} \right) s_\wedge + \frac{\left\| P_1- P_0 \right\|_1}{2} s_0.
\end{align}

Then, the concavity of $P\mapsto I_\alpha^{\texttt{a}, (t)}(P,\mathscr{W})$ given in item~\ref{I2-bb} together with Eq.~\eqref{eq:ec2} yield
\begin{align}
I_\alpha^{\texttt{a}, (t)}(P_0, \mathscr{W}) - I_\alpha^{\texttt{a}, (t)}(P_1, \mathscr{W}) &\leq H\left( \frac{\left\| P_1- P_0 \right\|_1}{2} \right) + \frac{\left\| P_1- P_0 \right\|_1}{2} \left( I_\alpha^{\texttt{a}, (t)}(s_0,\mathscr{W})  - I_\alpha^{\texttt{a}, (t)}(s_1, \mathscr{W}) \right) \\
&\leq H\left( \frac{\left\| P_1- P_0 \right\|_1}{2} \right) + \frac{\left\| P_1- P_0 \right\|_1}{2}  I_\alpha^{\texttt{a}, (t)}(s_0,\mathscr{W})
\end{align}
for $\alpha \geq 0$.
Thus, using the monotone increases of  $\alpha \mapsto I_\alpha^{\texttt{a},(t)}$ given in \ref{I2-aa} and recalling the definition of R\'enyi capacity given in Eq.~\eqref{eq:radius},
\begin{align}
\left| I_\alpha^{\texttt{a}, (t)}(P_0, \mathscr{W}) - I_\alpha^{\texttt{a}, (t)}(P_1, \mathscr{W}) \right| &\leq H\left( \frac{\left\| P_1- P_0 \right\|_1}{2} \right) + \frac{\left\| P_1 - P_0 \right\|_1}{2} C_{\eta, \mathscr{W}}.
\end{align}
The above inequality implies equicontinuity as desired.
\end{proof}

\bibliographystyle{myIEEEtran}
\bibliography{reference}

\end{document}